\documentclass[aps,prl,twocolumn]{revtex4-2}

\usepackage{blindtext,amssymb,gensymb,siunitx,textgreek,textcomp,mathtools,bm,soul}
\usepackage{wasysym} % needs to be loaded after ams package(s)!
\usepackage[version=3]{mhchem} % Formula subscripts using \ce{}

\usepackage[utf8]{inputenc}
\usepackage[T1]{fontenc}
\usepackage{subfiles}
\usepackage{fancyvrb}
\usepackage{lipsum}
\usepackage{enumitem}
\usepackage [english]{babel}
\usepackage [autostyle, english = american]{csquotes}
\MakeOuterQuote{"}

\usepackage[margin=0.625in]{geometry}
\usepackage{afterpage}
\usepackage{float}

\usepackage{caption,graphicx}
\usepackage{tablefootnote}

\usepackage{booktabs}
\usepackage[table,xcdraw]{xcolor}
\usepackage[hidelinks]{hyperref}
\usepackage{cleveref,xr}
\begin{document}

\title{From design to device: challenges and opportunities in computational discovery of p-type transparent conductors}

\author{Rachel Woods-Robinson*\textsuperscript{1,2,3}, Monica Morales-Masis\textsuperscript{4}, Geoffroy Hautier\textsuperscript{5}, Andrea Crovetto\textsuperscript{6}}

\affiliation{\textsuperscript{1}Clean Energy Institute, University of Washington, Seattle, WA, 98195 United States, \textsuperscript{2}Materials Sciences Division, Lawrence Berkeley National Laboratory, Berkeley, CA, 94720 United States, \textsuperscript{3}Materials Science Center, National Renewable Energy Laboratory, Golden, CO, 80401 United States, \textsuperscript{4}MESA+ Institute for Nanotechnology, University of Twente, 7500 AE Enschede, the Netherlands, \textsuperscript{5}Thayer School of Engineering, Dartmouth College, 14 Engineering Dr, Hanover, NH, USA, \textsuperscript{6}Centre for Nano Fabrication and Characterization (DTU Nanolab), Technical University of Denmark, 2800 Kongens Lyngby, Denmark}

\date{\today}

\begin{abstract}

A high-performance p-type transparent conductor (TC) does not yet exist, but could lead to advances in a wide range of optoelectronic applications and enable new architectures for, e.g., next-generation photovoltaic (PV) devices. High-throughput computational material screenings have been a promising approach to filter databases and identify new p-type TC candidates, and some of these predictions have been experimentally validated. However, most of these predicted candidates do not have experimentally-achieved properties on par with n-type TCs used in solar cells, and therefore have not yet been used in commercial devices. Thus, there is still a significant divide between transforming predictions into results that are actually achievable in the lab, and an even greater lag in scaling predicted materials into functional devices. In this perspective, we outline some of the major disconnects in this materials discovery process --- from scaling computational predictions into synthesizable crystals and thin films in the laboratory, to scaling lab-grown films into real-world solar devices --- and share insights to inform future strategies for TC discovery and design.

\end{abstract}

\maketitle

\section{Introduction}

Many advances in renewable energy technology are limited by the quality and availability of materials. Conversely, materials advances enable technological advances. The search for p-type transparent conductors (TCs) exemplify this: achieving a p-type TC with comparable properties to n-type transparent conducting oxides (TCOs) could enable new solar cell architectures and transparent electronics applications, serving as hole-selective top contacts or electrodes, buffer layers, back contacts for bifacial solar cells, and window layers in tandem devices.\cite{ginley2011transparent, willis2021latest} Many candidate p-type TCs have been proposed and explored, such as delafossites \ce{Cu$M$O2}\cite{nagarajan2001p} and \ce{CuI},\cite{wang2011native} among others. Yet the realization of high-performing p-type TCs is still a major research challenge, and p-type TCs for photovoltaic (PV) applications remain limited. This is in part due to fundamental physical trade-offs of combining transparency, p-type doping, and high hole mobilities in wide band gap oxides.\cite{delahoy2005transparent}
% This is due primarily to (1) intrinsic limitations in dispersion that localize the holes in the valence band, and (2) challenges in introducing free hole carriers and minimizing compensating defects. 

While the first p-type TCs have been found by serendipity and following chemical principles (e.g., Cu-based TCs), it is now possible to identify new p-type TCs using first principles computations and high-throughput screening. %Computational materials screenings have provided an approach to address these fundamental challenges in the pursuit of promising p-type TCs.
\nolinebreak "High-throughput" computational screenings usually begin with selection of a set of input compounds often from material databases (e.g., Materials Project\cite{jain2013commentary}, AFLOW\cite{curtarolo2012aflowlib}, OQMD\cite{kirklin2015open}, among others), followed by a series of filtration steps using computed descriptors which represent proxies for certain materials properties.\cite{brunin2019high}

As depicted in \autoref{fig:predictions}(a), after selection of inputs a typical screening method for p-type TCs starts with (1) a proxy for thermodynamic stability (usually energy above convex hull, $E_\mathrm{hull}$), (2) a proxy for absorption edge above the visible regime (usually the Kohn-Sham band gap, $E_\mathrm{G}$), (3) a proxy for high hole mobility (usually hole effective mass, $m_\mathrm{h}^*$), followed by a series of more computationally demanding calculations for hole dopability, band alignment, synthesizability, and other properties.\cite{hautier2013identification, varley2017highthroughput, woods2018assessing, brunin2019transparent} As final screening steps as shown in \autoref{fig:predictions}(b), higher accuracy computational methods can be applied such as hybrid or GW computations (to better estimate transparency) or electron-phonon limited mobility (to estimate hole transport).

Computational screening research from the past decade has yielded promising p-type TC candidates, and key discoveries are summarized in \autoref{fig:predictions}. One of the earliest studies focused on oxides and identified a handful new p-type TCO candidates.\cite{hautier2013identification} Since then, a wide array of follow-up research has refined the screening criteria or input material set, for example by extending the chemistries of interest beyond ternary oxides or by focusing on specific chemistries or structures.\cite{peng2013li, bhatia2015highmobility, sarmadian2016easily, williamson2016engineering, varley2017highthroughput, shi2017highthroughput, raghupathy2018rational, kormath2018database, yim2018computational, youn2019large, ha2019computationally, hu2020first}. These studies have motivated the synthesis and characterization of a handful of computationally identified candidates, with a few notable success stories. For example, very high mobility and transparency have been observed experimentally in moderately p-type \ce{Ba2BiTaO6}\cite{bhatia2015highmobility, shi2022modulation} and \ce{TaIrGe}.\cite{yan2015design} Transparency and high p-type dopability has been confirmed experimentally in \ce{Cr2MnO4}, although with lower hole mobilities.\cite{peng2013li} Overviews of experimentally realized p-type TCs can be found in existing literature reviews and benchmarking studies.\cite{willis2021latest} These successes, some of which are highlighted in \autoref{fig:examples}, are encouraging and bring a new level of predictability for computational materials science.

However, experimental confirmation of certain predictions has been either incomplete or not up to the expectations of the prediction. Boron phosphide (\ce{BP}), for instance, has been identified as a promising p-type TC with previous encouraging experimental results, but recent results indicate a difficulty to reach expected performances, as described later in this paper.\cite{varley2017highthroughput, ha2020boron, crovetto2022boron} \ce{Ta2SnO6} showed promise yet could not be doped enough to measure hole conductivity.\cite{barone2022growth} Other predicted materials remain to be grown and evaluated, ideally in thin film form (e.g., \ce{K2Sn2O3}, \ce{B6O}, \ce{NaNbO2}, \ce{BeSiP2}).\cite{hautier2013identification, youn2019large, woods2023designing} Progress in the field of predicted n-type TCs serves as an indicator that advances in predicted p-type TCs are within reach; for example, as a success of a high-throughput screening, a new computationally predicted n-type TCO \ce{ZnSb2O6} has been recently confirmed by experiment with high doping and mobility.\cite{hautier2014does, jackson2022computational} However, despite the many successes, scaling up predicted TCs for efficient optoelectronic device applications remains a major challenge, and a variety of disconnects have emerged. 

%There is something to be said about CaTe and LiCuO as well. Same for LiNbO2

Historically, unwanted lab results and corresponding insights from researchers are usually not published, leading to a literature bias and a difficulty disentangling why disconnects are occurring. In this article, we identify and contextualize these disconnects observed in the literature and within our own research in scaling p-type TC predictions into PV devices. First, we design a framework for the different stages involved in this process, as summarized in \autoref{fig:framework}. Next, we highlight some of the most pernicious disconnects (see \autoref{fig:disconnects}). One set of disconnects stem from transforming predicted candidates into thin films in the lab: due to challenges including synthesizability, phase purity, dopability, and unwanted absorption, so far no predicted p-type TC grown in the lab performs on par with n-type TCOs. Another key set of disconnect arises from transforming predictions into scalable optoelectronic devices: challenges emerge such as making sufficient contact to the device, tuning band alignments, and growing interfaces free of defects, and other barriers. We also discuss specific barriers related to scalability and sustainability. Lastly, we offer guidance on how the research community can overcome such disconnects to bring our predictions into fruition in p-TCs and beyond.

%%%%%%%%%%%%%%%%%
% Design2device %
%%%%%%%%%%%%%%%%%

\section{Materials design-to-device framework for p-type TCs}

\subsection{PV applications of p-type TCs}

While conventional silicon (Si) solar modules do not typically include contact layers, emerging PV technologies such as silicon heterojunction (SHJ), tunnel oxide passivated contact (TOPCon), thin film (e.g., CdTe, perovskites, and other direct band gap absorbers), and tandem solar cells require various contact and buffer layer configurations to transfer charge carriers throughout the device. The current market share of PV (as of 2024) is mostly conventional Si, however various projections show that meeting our net zero requirements by 2050 will likely require rapid development and scaling of such emerging PV technology.\cite{vdma2024itrpv}

% \afterpage{
\begin{figure}[H]
    \centering
    \includegraphics[width=0.45\textwidth]{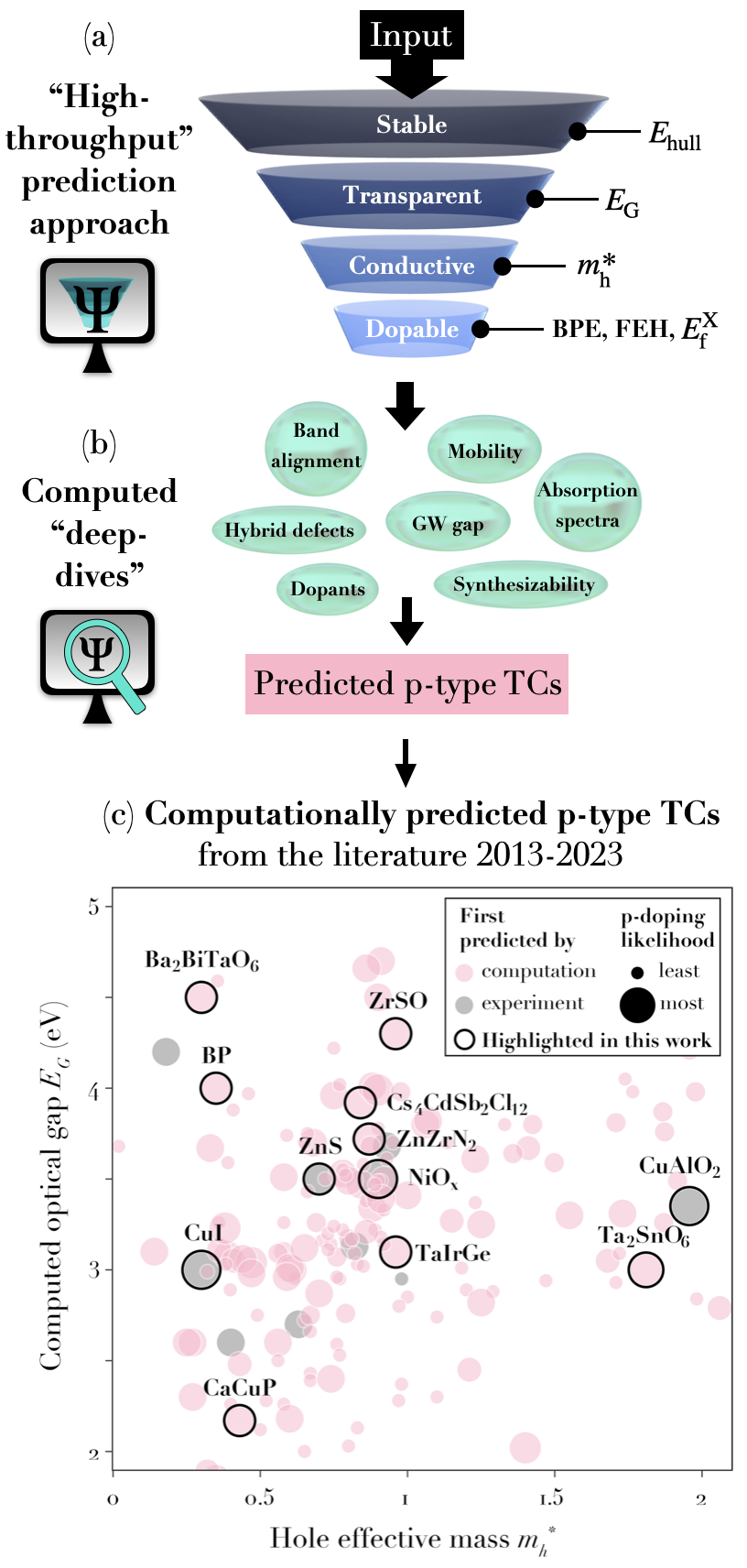}
    \caption{(a) Example high-throughput screening method for p-type TCs.\cite{woods2018assessing}. (b) Example of properties that are assessed during "deep dive" calculations. (c) A literature review of p-type TC candidates predicted from high-throughput computational studies, with candidates highlighted that are discussed in this article. Pink shaded materials have been predicted first as p-type TCs by computation, while grey shaded materials were first demonstrated experimentally and are included here as references. "Most" likely p-type dopable have been confirmed by experiment or have hybrid defect calculations, while "least" likely have not been assessed for dopability. A comprehensive table with references of computational studies of p-type TCs and a \texttt{jupyter} notebook to reproduce (c) are provided in the Supplemental Material.\cite{si}}
    \label{fig:predictions}
\end{figure}
% }

\noindent \nolinebreak One recent study projects deployment of these emerging PV technologies to ramp up rapidly after 2030, and by around 2040 overtake conventional silicon as the dominant PV technology.\cite{stanbery2023photovoltaic} However, as researchers develop these new configurations, few options for inorganic p-type transparent contacts are available (which generally have better conductivity and stability than organic hole transport materials), and this limits available architectures and resulting performance. Therefore, developing appropriate contact materials for emerging PV, especially p-type transparent contacts, is an important research direction.

\begin{figure*}
    \centering
    \includegraphics[width=\textwidth]{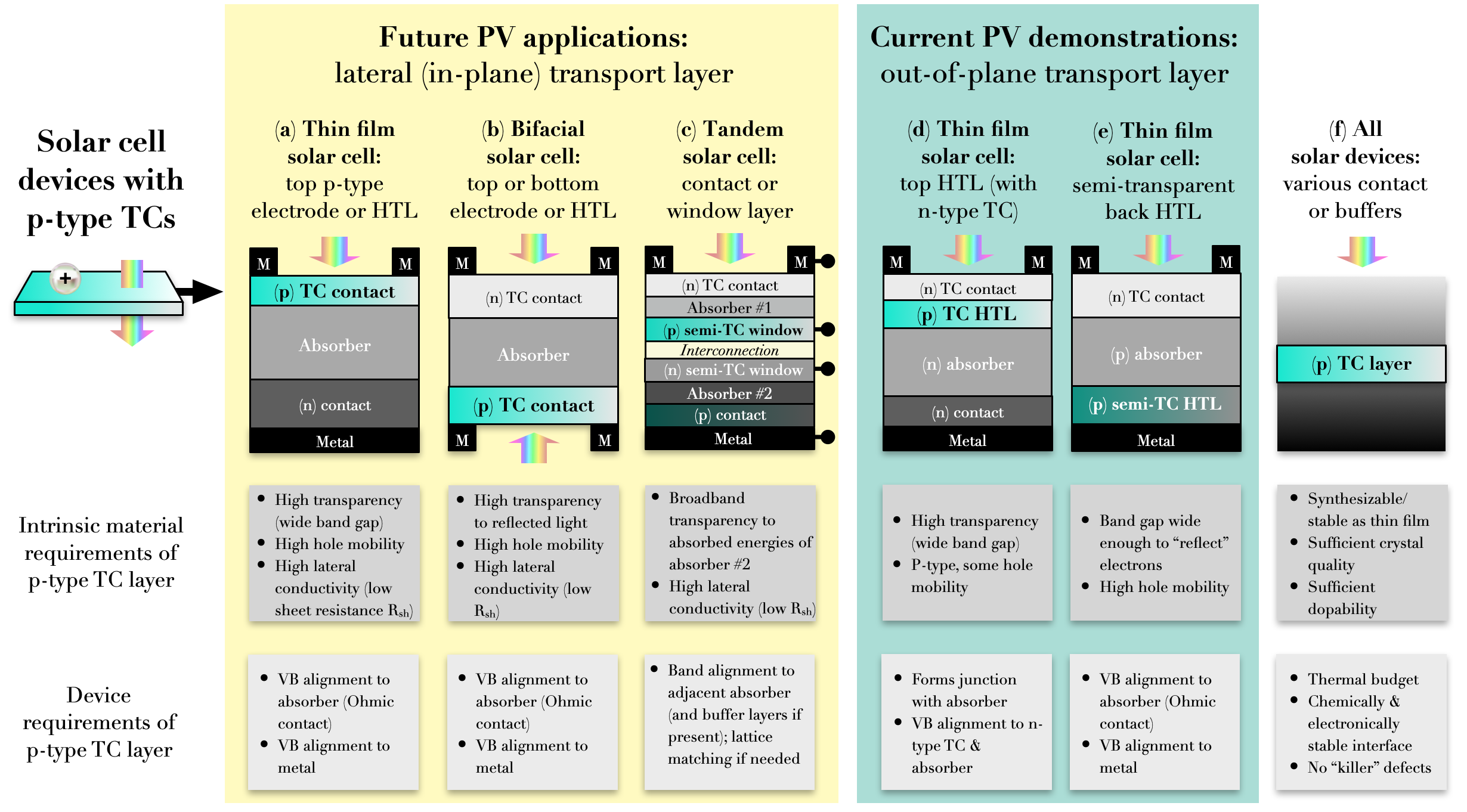}
    \caption{Schematic solar cell device architectures that incorporate p-type TCs. The focus of materials in this paper is to enable architectures for "Future PV applications," including as (a) a top p-type electrode to a thin film single junction stack, (b) a top or bottom electrode to a bifacial solar cell, and (c) as a contact or window layer to a tandem solar cell. Panel (c) depicts a 4 terminal mechanically-stacked tandem, in which a p-type semi-transparent conductor ("semi-TC") layer could serve as either the bottom contact of absorber \#1, as shown in (c), or the top contact of absorber \#2 if polarities are reversed. A p-type semi-TC could also be used in other tandem configurations such as a window layer in a 2T monolithic tandem, however in other cases lateral transport may be less important. Reported demonstrations of p-type TCs in PV, in which lateral transport is also less important, are included for reference: (d) semi-transparent bottom electrode or buffer for thin film solar, and (e) hole transport layers (HTLs) in contact with an n-type TC. Panel (f) summarizes properties that must be attained in each configuration. This figure is intended to represent the key functionalities that p-type TCs could enable, and is by no means exhaustive. Layers are simplified for clarity, and thickness is not to scale.}
    \label{fig:device-designs}
\end{figure*}

Applications for p-type TCs span far beyond solar (e.g., flat panel displays, transparent electronics, etc.), but for the purposes of this article we focus mainly on PV devices. Within the field of emerging PV, p-type TCs with different properties could serve a variety of roles. For example, p-type TCs could have a double role, that of a hole carrier selective contact (where a high work function is required) and that of a transparent electrode (where a low sheet resistance is required, usually below 100 $\Omega$/sq). As a selective contact, the p-type TC would extract photogenerated holes in the absorber material, and, as transparent electrode, it would laterally transport the carriers to the device terminal. Since the material is transparent, it would also function as a window layer in the device. This would enable novel device architectures for e.g., bifacial and semitransparent PV, which currently relies exclusively on the use of n-type ITO.

Representative device designs are summarized in \autoref{fig:device-designs}, with future applications enabled by high-performance p-type TCs highlighted in yellow in (a--c).  Designing TCs with high lateral (or "in-plane") transport for these applications will be the focus of this paper. For reference we also show two "out-of-plane" transport configurations (d-e) in which mobility and dopability are less important, buffer HTLs and semi-transparent contacts (in which transparency is less important); since these applications have actually been demonstrated, they will be used to showcase challenges scaling to devices in the "Lab-to-device disconnects" section, but are not the focus of this paper. This variety of applications is the origin of one important disconnect, which arises from an intermingling of problem statements during the materials design stage. Namely, a material predicted from an application-agnostic screening for p-type TCs may not be appropriate to use in a specific application, e.g., might not provide an adequate band alignment.

\subsection{Materials discovery stages}

\begin{figure*}
    \centering
    \includegraphics[width=\textwidth]{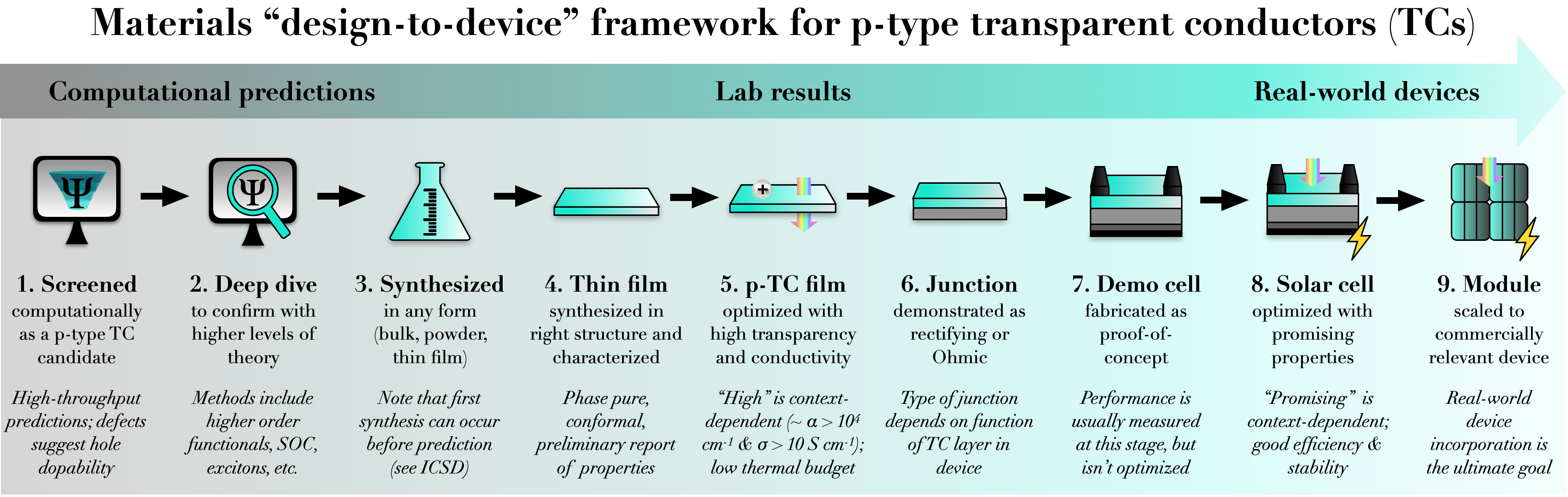}
    \caption{Our proposed framework of materials "design-to-device" progression for p-type transparent conductors (TCs), representing stages of technological innovation from computational prediction (far left) to a scalable device (far right). \autoref{fig:predictions} highlights computational methods within the first two stages of this process, while \autoref{fig:device-designs} outlines various devices designs in stages 7--9.}
    \label{fig:framework}
\end{figure*}

To understand the disconnects that emerge when scaling up computationally-predicted materials, we first summarize the main stages involved in going from a predicted material, to a synthesized material, to the exploration of its functionality on a device, and to a commercially-relevant device. For the specific case of a p-type transparent conductor (p-type TC), the "design-to-device" framework is summarized in \autoref{fig:framework} with the following nine key stages:

\begin{enumerate}[wide]

    \item \textbf{Computational screening}: As shown in \autoref{fig:predictions}(a) a material is "screened" as a promising candidate, often from a database. Simple computational metrics representing properties of stability, transparency, and conductivity are evaluated. Usually a stoichiometric, well-defined crystal structure is assumed, though high-throughput defects have started to emerge and could be used at this stage.\cite{dahliah2021high, broberg2023high} Various biases can occur at this early stage that introduce disconnects at later stages.
    \item \textbf{Computational deep dive}: Higher levels of theory (e.g., spin-orbit coupling, GW) and more in-depth analysis methods are applied to more accurately predict properties of synthesizability, transparency, temperature-dependent mobility, carrier concentration from defects, and device-relevant parameters such as band alignment, as shown in \autoref{fig:predictions}(b). Structural models can be more complex at this stage to account for off-stoichiometric and non-equilibrium effects.
    \item \textbf{First experimental realization}: The material is synthesized for the first time, usually in the form of a bulk single crystal or powders (e.g., using solid state synthesis or mechano-chemical synthesis). From powders or single crystals it is possible to confirm crystal structure, stoichiometry, air and moisture stability, and preliminary properties, though at this stage optoelectronic properties are not usually assessed. When screening from a database of known materials such as ICSD, this stage has often already been realized, so occurs before stage 1.
    % However, assessing device-relevant properties requires that the compound is fabricated in the thin film form.
    \item \textbf{Thin film synthesis}: The material is synthesized as a phase-pure, conformal thin film. There are numerous methods to fabricate thin films, from wet synthesis to chemical or physical vapor deposition, in one or multiple steps. Thin film formation can result in epitaxial (single crystal), polycrystalline or amorphous films. The microstructure, thickness, stoichiometry, point or crystallographic defects, and surface effects will strongly influence the thin film properties and functionality. Specific synthesizability challenges and disconnects for p-type TC thin films have been discussed previously by Fioretti and Morales-Masis. \cite{fioretti2020bridging}
    % All this creating one of the critical disconnects between the ideal material properties and the measured thin film properties. 
    \item \textbf{Thin film optimization}: A p-type TC film is optimized, with high transparency and conductivity (the first reported film in stage 4 is usually not indicative of how much performance can be tuned). Such optimization can occur through doping, alloying, or optimizing process conditions such as annealing, among other methods. Throughout this manuscript, "high" transparency refers to $\alpha > 10^4$ cm\textsuperscript{-1} and "high" conductivity refers to $\sigma > 10$ S cm\textsuperscript{-1}, though what is considered sufficiently "high" depends on application. We note this stage is often skipped for novel p-type TCs.
    \item \textbf{Junction demonstration}: A p-type TC is demonstrated to form a junction with another material, either deposited directly on an active substrate or on top of a device stack. The key challenges at this stage are chemical and thermal compatibility (e.g., to avoid chemical intermixing at interfaces or material degradation) and electronic alignment (e.g., valence band energies aligned), as well as good physical adhesion, without degrading the junction partner material.
    % At this stage, there is a disconnect between growing the thin film under ideal conditions (crystallization temperature, lattice-match film growth) and the device conditions (e.g., distinct bottom layers influencing the growth of the p-type TC, limited temperatures resulting in non-crystalline films).
    \item \textbf{Solar cell demonstration}: The p-type TC film is incorporated into a full solar cell device stack with demonstrated performance. Often this first device has low efficiency and poor device characteristics. p-type TCs can be deposited directly on a substrate or on top of the device stack (depending on the device polarity). The substrate or device stack will define the limitations for the p-type TC film deposition, such as thermal budget. This in turn will influence (or limit) the final properties of the p-type TC film.
    \item \textbf{Solar cell optimization}: The solar cell is optimized around the p-type TC layer, with reasonable PV performance values such as open circuit voltage, fill factor, and efficiency. At this stage, in addition to retaining good thin film properties and optimizing the TC, the interfaces play a key role in device performance. One way this disconnect can be alleviated by including interface simulations once the device stack is known (e.g., as done for organic and halide perovskite solar cells).\cite{minemoto2001theoretical}
    \item \textbf{Scaling device to module level}: Scaling up lab-scale devices (solar cells) into commercially-relevant devices (solar modules) requires different fabrication techniques for TCs, and can change properties and requirements. As reported in NREL's well-known PV efficiency charts,\cite{nrel2023best, nrel2023champion} solar cell properties often cannot be reproduced on the module scale, due to various disconnects such as conformality and processing constraints. At this stage, supply chain, cost, and sustainability factors also must all be considered. The ultimate goal of materials discovery is to go from a predicted material to this final stage.

\end{enumerate}

This framework will be used throughout this manuscript to guide our exploration of disconnects. These steps are inspired by the technology readiness level (TRL) framework, which will also be referred to when appropriate, however our steps 1--6 expand upon earlier stages of materials design. This framework is designed and adopted specifically to address stages of materials discovery in p-type TCs. We emphasize that discovery does not necessarily follow this linear flow: often, optimization involves iteration between these stages. Historically, most materials have been synthesized first before computational predictions, and many recent studies return to computational "deep dives" after experimental stages to fill in experimental knowledge gaps or inform decisions. To highlight representative materials discussed throughout this paper, \autoref{fig:examples} illustrates the stage along the "design-to-device" framework in which research and development is stalled, a graphical highlight of research at this stage, and one of the key disconnects describing why progress is stalled.

% Under operation, more challenges can arise as interfacial defect energy levels change.

%%%%%%%%%%%%%%%%%
%% Comp to Lab %%
%%%%%%%%%%%%%%%%%

% \section{Prediction disconnects}

\section{Design-to-lab disconnects}

\begin{figure*}
    \centering
    \includegraphics[width=180mm]{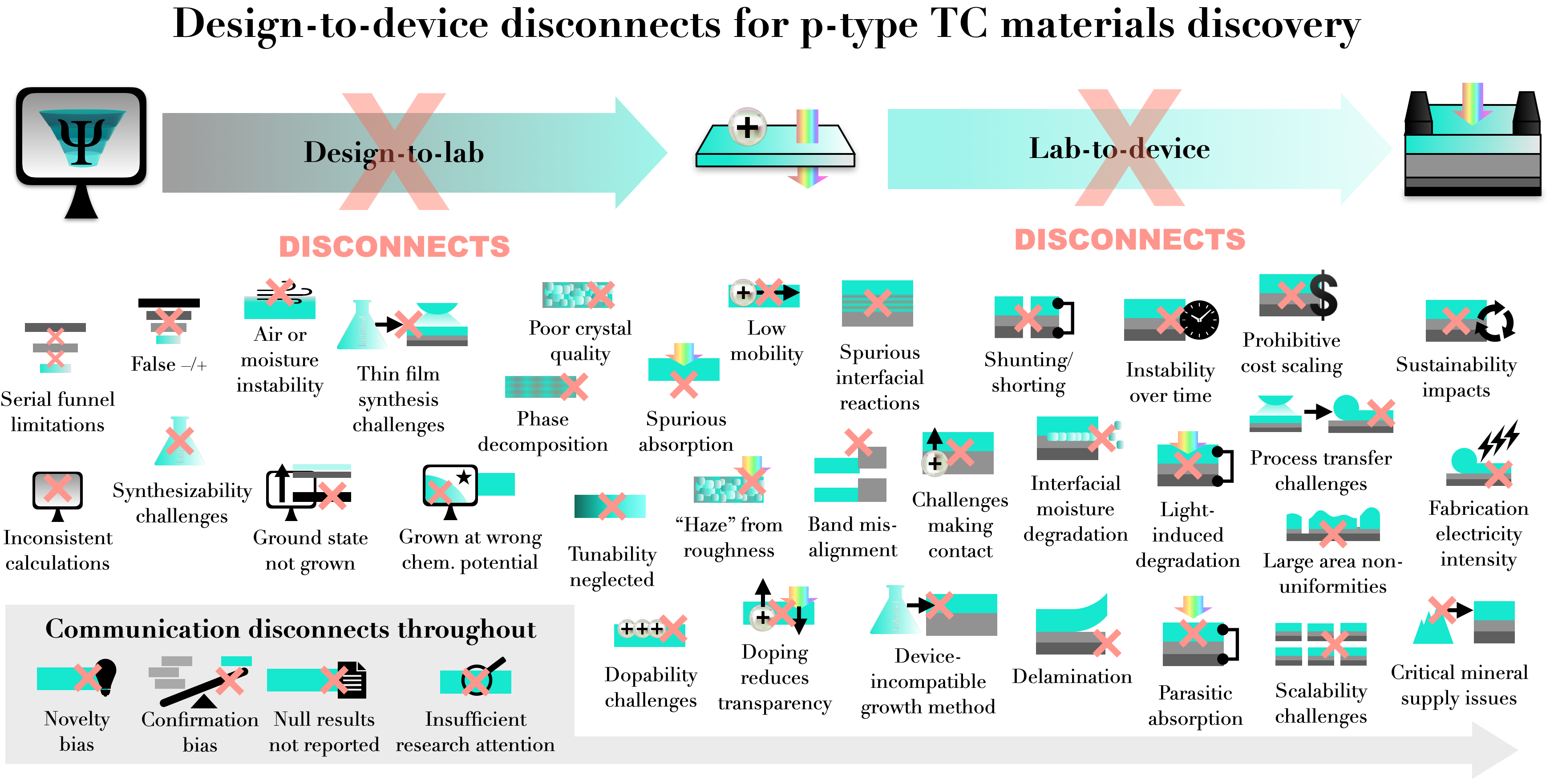}
    \caption{A schematic summary of the primary disconnects in scaling computationally predicted p-type TCs into solar cell devices.}
    \label{fig:disconnects}
\end{figure*}

In this section, we outline key disconnects that emerge going from the prediction of a new p-type TC (stage 1) to the realization of optimized properties in an experimental p-type TC thin film (stage 5). Some common disconnects at this stage are summarized in the left-hand side of \autoref{fig:disconnects}. We will also discuss strategies to confront these challenges.

We note that "figures of merit" (FoMs) have been used to assess and optimize TCs. However, there have been multiple FoMs proposed (Fraser and Cook,\cite{fraser1972highly} Haacke,\cite{haacke1976new} Gordon,\cite{gordon2000criteria} Haacke High Resolution,\cite{cisneros2019resolution} among others) and no consensus in the community which is best, in particular for p-type TCs. The selection of an appropriate FoM is highly application-dependent and material-dependent, with potential for misuse in material selection based on FoM alone. We caution against using a single, generalized, application-agnostic FoM, recommending instead to either report optical and electrical measurements separately or define an FoM for a specific application, and therefore we do not report FoMs (see instead reported FoMs from experimental literature\cite{zhang2016p,mirza2023role,geng2023amorphous}).

\subsection{Computational screenings contain limitations and biases}

First principles computations are powerful but have inherent fundamental limits and numerical dangers such as convergence that need to be kept in mind. We will assume in this discussion that computations have been performed by an expert, achieving the necessary numerical convergence for the targeted task. We will also assume that the inherent limits of the approximation used (e.g., semilocal DFT underestimates band gap and affects defect computations dramatically) are taken into account in the screening process. Unfortunately, a small fraction of studies in the field do not follow these requirements and we warn the reader to always question whether computational results have been interpreted with caution and the right amount of expertise. 

The efficacy of high-throughput computational screenings emerges from the quality of its input data, which usually comes from computational databases or the ICSD. These repositories have grown substantially, encompassing diverse materials and properties. However, extensively studied materials are over-represented in these databases, which can induce a selection bias toward well-studied candidates and can potentially neglect less-studied ones. For example, previous literature suggests non-oxides as promising p-type TCs due to low effective masses and low ionization potentials.\cite{varley2017highthroughput} However, there are far more oxides than non-oxides in materials databases, so non-oxides are under-sampled in screening studies.

An algorithmic bias can lead to prioritizing properties that are more readily calculated or accessible, potentially overlooking properties that may be more representative of desired performance. In order to reduce the input materials to a manageable subset, the initial steps of computational screenings usually filter with property proxies that are computationally inexpensive ("cheap") to calculate. For example, although both effective mass $m^*$ and scattering time $\tau$ contribute to achieving high mobility, given current algorithms $m^*$ ("cheap" proxy) is much easier to calculate than $\tau$ ("expensive" proxy). Additionally, hole conductivity depends on doping density which is also difficult to compute. Therefore, we are biased towards looking for materials with low $m^*$ rather than those with excellent scattering or doping properties, and we need to be careful about how close a cheap proxy correlates to an expensive proxy.
% Additionally, databases include ordered materials because they're easier to calculate, while disordered off-stoichiometric materials are more representative of real p-type TCs. 

Similarly, a knowledge bias occurs when a model or analysis is limited by existing knowledge and fails to account for emerging or unexplored factors. For example, screening for low effective mass and wide direct band gap as shown in \autoref{fig:predictions}(a) is appropriate for traditional TCOs that behave as doped semiconductors following a rigid band model (e.g., \ce{ZnO}, \ce{SnO2}). More recently proposed mechanisms such as polaronic TCs,\cite{brunin2019high} intrinsic TCs\cite{zhang2015intrinsic}, or TCs containing forbidden transitions\cite{woods2023designing} are less likely emerge from this approach, however this can be remedied when new mechanisms are understood and can be incorporated into the screening.

Once screening criteria are chosen, a threshold bias can arise from selecting cutoffs, their tolerances, and criteria priorities, as these thresholds often lack objectivity. For instance, picking a cutoff for "low" hole effective mass ($m^*_\mathrm{h}$) is typically arbitrary; opting for only the lowest $m^*_\mathrm{h}$ could exclude highly transparent materials, and the $m^*_\mathrm{h}$ needed to achieve a certain high mobility depends on material physics. Additionally, there is a lack of threshold standardization regarding how defect calculations are performed and interpreted. While serial screening is common, Pareto analysis or multi-objective optimization can counter threshold bias and improve outcomes.\cite{lejaeghere2013ranking} However, one key challenge is selecting appropriate weighting parameters formally and quantitatively. Such optimizations become application dependent and are rarely done for p-type TCs.

Lastly, biases can occur when reporting computational results.  Novelty biases can lead to publishing and highlighting new and exciting findings, while negative or non-novel outcomes are not usually reported. Screening studies tend to report "new" p-type TC predictions rather than reporting whether past predictions have been corroborated. However, especially given challenges to synthesizing predictions in the lab (see next sections), it would be extremely useful for experimental researchers to know whether a result has been computationally reproduced by multiple groups and different methods. One approach could be to establish an open computational database specifically for p-type TCs to enable comparison of calculation parameters and results; such an effort would require dedicated funding and staff for development and maintenance.

Together, these biases can lead to false negatives and false positives emerging from screenings. Input sets and screening criteria need to be judiciously selected and appropriately communicated to avoid misleading predictions. Although false negatives are challenging to address, false positives can be mitigated by performing a computational "deep-dive" (stage 2) before attempting synthesis. As an example from our research, shown in \autoref{fig:examples}, sputter synthesis was attempted (stage 3) shortly after \ce{ZnZrN2} emerged from a p-type TC screening (stage 1), before a thorough computational assessment (stage 2). The target phase could not be synthesized, and by turning back to a "deep dive" it was hypothesized that due to disorder intolerance the selected synthesis method would likely have never yielded this phase.\cite{woods2022role} In general, experimentalists using screenings to inform their research should be aware of how such biases propagate and how to carefully scrutinize the extent of computational findings before going to the effort to grow it in the lab.

\subsection{Predicting synthesizability and stability can be challenging}

\subsubsection{Bulk crystal synthesis}

% After a screening, more accurate synthesizability metrics can be applied to the smaller subset of outputs, especially for compounds that have not yet been synthesized before as thin films.

The next disconnect is that many new p-type TCs have been predicted, but not all have been successfully synthesized in the laboratory in the desired crystal structure (stage 3 in \autoref{fig:framework}). One likely reason for the lack of reported synthesis is that many have not been attempted yet, or if attempted have not been published possibly due to negative results (so-called "dark reactions"\cite{gautier2015prediction}). Computational databases tend to report whether a given material has been synthesized previously (pointing to the ICSD), and bulk or powder synthesis is most common. One approach to avoid assessing synthesizability is to simply restrict input datasets to just these known experimental materials. However, some exciting new predictions of p-type TCs do not yet have any reports of synthesis, e.g., \ce{Cs$MCh$2} ($Ch$=S,Se,Te)\cite{shi2017highthroughput} and double perovskites such as \ce{Cs4$M$^{2+}$B$^{3+}2$X$^{VII}12},\cite{xu2018prediction} so for these materials synthesizability should be judiciously assessed. We will start by discussing challenges that have arisen with growing hypothetical crystals that have not yet been synthesized in any form.
% As discussed previously, in our research we have been unable to synthesize several predicted TCs in their targeted phase, stalling research in stage 1--2 of \autoref{fig:examples}.

Accurate assessment of whether a hypothetical material is actually synthesizable and stable is not trivial. The first disconnect arises from the crude assumption that the descriptor $E_\mathrm{hull}$ is a good proxy for synthesizability. Namely, it is generally assumed that materials with formation enthalpies at or near the ground state energy predicted by DFT at zero Kelvin ($E_\mathrm{hull} \simeq$ 0 eV/atom) are the most likely to be experimentally realized, while materials with  $E_\mathrm{hull} \gg$ 0 eV/atom are less likely to be synthesized.\cite{sun2016thermodynamic, aykol2018thermodynamic, hegde2020phase} $E_\mathrm{hull}$ itself is impacted by errors in DFT and correction schemes, and does not include entropic effects. Various low-cost approaches to estimate vibrational and configurational entropy in computational screenings have emerged,\cite{bartel2018physical, stevanovic2016sampling} which can be incorporated screenings. Additionally, assessing dynamic stability is essential,\cite{haastrup2018computational} and synthesis is driven by kinetics as well as thermodynamics. Despite many recent first-principles and machine learning (ML) efforts to address kinetics via synthesis pathways,\cite{aykol2021rational, szymanski2021toward, davariashtiyani2021predicting, huo2022machine, szymanski2023autonomous, montoya2024ai} such as solid-state reaction networks,\cite{mcdermott2021graph, mcdermott2023assessing} a deeper understanding and predictability is still needed. Increasing efforts in automating solid state synthesis may help address this challenge, through solution-based approaches such as robotic arms to handle chemical synthesis as well as various high-throughput vacuum-based approaches.\cite{szymanski2021toward, yano2022case, gregoire2023combinatorial}
To our knowledge autonomous laboratories have not yet been tailored towards p-type material discovery, but could be a promising research direction especially for exploring processing conditions to promote phase stability, and optimizing dopants and off-stoichiometries.

Specific synthesizability challenges can arise due to the complexity of compounds selected as p-type TC candidates.\cite{toher2019unavoidable} Many predicted TCs are multinary materials, since multiple elements provide an increased search space and tunability.\cite{walsh2020holey} For instance, \ce{CuAlO2} (often considered the first p-type TC) can be considered a ternary counterpart to p-type \ce{Cu2O} in which Al increases the band gap,\cite{kawazoe1997p} and similarly \ce{Ba2BiTaO6} a quaternary extension of p-type \ce{BaBiO3} in which Ta increases the gap. Furthermore, binary TCs can be extended into multinary solid solutions to enhance performance, such as recent S incorporation into CuI.\cite{ahn2022highly, geng2023amorphous, mirza2023role} However, more elements can induce challenges in synthesizability including increased likelihood to phase segregate, as well as complex disorder and defect configurations.\cite{toher2019unavoidable, hegde2020phase} For example, one challenge that has stalled progress in thin film synthesis of predicted p-type TC \ce{ZrOS}\cite{hautier2013identification} has been phase separation into defective binary phases \ce{ZrO_x} and \ce{ZrS_x}, as it is challenging to introduce S into \ce{ZrO_x} (see \autoref{fig:examples}).\cite{fioretti2020bridging} Another well-known example of such challenges can be observed in quaternary and quinternary kesterite PV absorbers, which tend to easily phase separate.\cite{kumar2015strategic}

Even if reported as "synthesized" in the ICSD, many compounds are not stable in air or moisture. For example, \ce{K2Sn2O3} was predicted as a p-type TC by Hautier et al.\cite{hautier2013identification}, but rapidly degrades in air. Air instability does not completely rule out a candidate, but using such materials in a device and even measuring them introduces serious challenges. For example, SnO and AgI are unstable in air, but have still been explored for device applications. CuI can also present stability challenges when exposed to air and moisture, presenting variations in conductivity due to iodine loss or Cu-oxidation. However, this depends strongly on the fabrication method and solar cell process used.   \cite{liang2010improvement, cha2017air, crovetto2020water} Non-oxides in particular tend to be less stable in air and moisture; some are protected by a passivation layer, while others can fully decompose or oxidize. One simple thermodynamic check for moisture instability is the descriptor "energy above the Pourbaix hull" ($E_\mathrm{pbx}$), based on Pourbaix diagrams.\cite{singh2017electrochemical} Another recent method uses a "greedy algorithm" to assign heats of oxidation to screen for in-oxygen environmental stability and heuristically determine whether a material may self-passivate.\cite{twyman2022environmental} However, to our knowledge there is no extensive benchmarking to show that that these metrics are predictive of experimental instabilities. In our experience, especially with non-oxide TCs, p-type TC candidates predicted stable by $E_\mathrm{hull}$ and even $E_\mathrm{pbx}$ have still degraded in air. Air instability may be a key reason why progress in non-oxide p-type TCs has been limited.

% A very good example is in thermoelectric. Yb14MnSb11 is not air stable thermodynamically but used at 1200K without issues... This is very difficult to predict nowadays.

\subsubsection{Thin film synthesis}

Specifically for TCs, another set of disconnects arises going from bulk synthesis to thin film synthesis (stage 3-4). Computational predictions are generally derived from bulk, periodic calculations at thermodynamic equilibrium, whereas in practical applications p-type TC materials are thin films, contain multiple types of defects, and are usually synthesized using non-equilibrium growth techniques. Sometimes these assumptions hold and, for example, a predicted structure can be grown as a thin film with properties corresponding to predictions, as demonstrated by various success stories such as \ce{Ba2BiTaO6} and \ce{CaCuP}.\cite{bhatia2015highmobility, willis2022prediction} However, this is not always the case; the effects of surface energies, surface termination, strain, and texturing in thin films can influence which phase is stabilized at a given condition, and can yield a range of properties far beyond the single observable value predicted for the bulk structure. Additionally, amorphous thin films are difficult to simulate, though many high-performing n-type TCs are amorphous.\cite{aykol2018thermodynamic}

Similarly, different thin film synthesis methods can result in an extensive span of structural properties that can deviate from predicted bulk properties. However, there is not yet a framework such that an ideal synthesis route and tool can be selected from first principles alone. The synthesis pathway prediction algorithms mentioned previously are a step in this direction, but these assume thermodynamic equilibrium and are method agnostic (many polycrystalline thin film synthesis methods are non-equilibrium). From the experimental side, the synthesizability challenge is to grow thin films that can tolerate defects in microstructure, surface effects, grain boundaries, strain, dislocations, etc., while retaining properties simulated from single crystalline materials (see next section).\cite{woods2018assessing} In sputtering, a technique commonly used for exploratory synthesis as well as for commercial TCOs, in some cases the "effective temperature" at the surface of a growing film have been shown to be more predictive of synthesizability and structural properties, rather than the chamber temperature.\cite{woods2022role} One disconnect here is simply that in stages 3 and 4, not all synthesis pathways have been sampled. Confirmation biases can arise if researchers try one approach, encounter difficulties, and then decide to abandon the material altogether without exploring alternative approaches to access a different region of configuration space.

% "Stability" is often lumped together with "synthesizability", often also approximated using $E_\mathrm{hull}$, but in practice whether a material is synthesizable or stable is determined by different physical factors. Computed thermodynamic stability can be a useful metric to determine whether a material is likely to phase segregate over time, however other aspects of stability (such as ???) also need to be assessed for real applications. 

\subsection{Origin of the mismatches between theoretical and experimental optoelectronic properties}

If a predicted p-type TC can be successfully synthesized as a thin film (stage 4), the next challenge (stage 5) is to experimentally achieve predicted properties of interest such as high optical transparency and high hole mobility and concentration. Yet in some cases of predicted p-type TCs grown in the lab, at least one of these properties falls short of the computationally predicted ideal. There may be three reasons for these disconnects: (1) limitations of high-throughput descriptors in accurately estimating the optoelectronic properties, (2) differences in material properties between an ideal bulk crystalline material (the object of computational modeling) and a practically realizable thin-film material, and (3) extrinsic experimental issues negatively influencing the predicted properties of a thin-film material.

\subsubsection{Accuracy of high-throughput optoelectronic properties}

First-principles calculations are useful to approximate experimental properties, but they also come with caveats. In general more "expensive" calculations improve accuracy, but there are still limits and trade-offs. Rather than review computational methods to screen for p-type TCs, which has been done in depth elsewhere,\cite{woods2018assessing, brunin2019transparent, cao2019design, willis2021latest} here we will highlight a few key limitations to computing transparency, mobility, and dopability in bulk crystalline materials. We note, however, that the most efficient high-throughput approaches are built using a tiered screening approach where cheap methods are used first to select materials (taking into account their lower accuracy) and more accurate and expensive methods are used for the final candidates before moving up to experimental verification (see \autoref{fig:predictions}).

% [Transparency]
A wide band gap is a central property of a TC. While "traditional" semilocal DFT is known to significantly underestimate band gaps (sometimes by 1 to 2 eV),\cite{chan2010efficient} many more accurate methods with different levels of costs are available nowadays such as GW or hybrid functionals.\cite{brunin2019high} Some of these methods have already been used in a high-throughput fashion and machine learning is speeding up the assessment of high accuracy band gaps with minimal computational cost.\cite{van2017automation, chen2021learning, dahliah2021high, yuan2024discovery} Beyond the accuracy of the method, one needs to keep in mind optical gaps are not electronic gaps. A better estimate of the optical gap can be obtained using relatively cheap independent particle approximation (IPA) computations on top of DFT or hybrids. These computations will take into account the transition dipole moment for optical absorption (i.e., the oscillator strength) with possible forbidden or weak transitions. Some important TCs such as \ce{In2O3} rely on forbidden transitions for their high transparency.\cite{walsh2008nature} Recent high-throughput work including in the TC field has started including these optical absorption computations, and p-type TCs with forbidden transitions have recently been proposed.\cite{woods2023designing} We note that more expensive approaches beyond IPA such as TD-DFT or Bethe-Salpeter equations (BSE) can be used to provide more accurate optical absorption, but in general, TCs with a band-like conduction mechanism have effective masses and band gaps that lead to small excitonic effects (exciton binding energies of around the tens of meV).\cite{baranowski2020exciton}

To model transport, effective masses ($m^*$) that can be obtained from DFT or higher order theories are most commonly used as a proxy for carrier mobility. While effective masses vary between semilocal DFT and more accurate methods, the difference is usually small enough for semilocal approaches to suffice, especially when using $m^*$ for screening and as a proxy for mobility. While $m^*$ is a simple concept for conventional single-parabolic band materials, real materials can be anisotropic and have complex band structures with several competing bands.\cite{brunin2019high} Discrepancies between computed values of $m^*$ for the same material in the literature often come from different crystallographic directions or bands. When used as a proxy for mobility, it is important to consider all the possible bands involved in transport and average them out. Otherwise, picking the highest curvature band tends to underestimate the effective mass. In the last five years, the full computation of mobility has made enormous progress. Codes such as \texttt{Perturbo},\cite{zhou2021perturbo} \texttt{EPW}\cite{lee2023electron}  and \texttt{ABINIT}\cite{brunin2020phonon} have well-documented and tested open source codes that fully compute phonon-limited mobility. These comprehensive approaches have been even scaled up and can be used to compute hundreds of materials automatically.\cite{claes2022assessing} They remain expensive however, and simpler approaches assuming simplified scattering processes have emerged such as \texttt{AMSET}, which also provides a high quality open source code.\cite{ganose2021efficient} Errors of around 20\% compared to experimental mobilities have been reported for these approaches based on electron-phonon and Boltzmann transport theory.\cite{ganose2021efficient} The field of TC design has already started to use these methods in recent studies. \cite{jackson2022computational, willis2022prediction, woods2023designing, willis2023limits, ha2019computationally} 

Finally, defects are essential to understand and predict p-type TCs. Point defects control doping and intrinsic defects can prevent p-type doping. Certain defects (e.g., oxygen vacancies) can act as "hole-killers" and prevent p-type doping. Due to the lack of automation and high computational cost, early high-throughput studies only computed defects as a "deep-dive" step after low $m^*_\mathrm{h}$ and wide band gap materials had been identified.\cite{hautier2013identification} However, since then automation has made huge progress and software packages have emerged such as \texttt{pycdt} and \texttt{pylada} that facilitate the generation and processing of input and output files (including charge corrections).\cite{broberg2018pycdt, goyal2017computational} 

Nowadays, key challenges stem not from automation but rather from the compromise between expensive, large supercell computations with hybrid functionals (which is the gold standard for defect computations) and the cheaper semilocal DFT-based approaches. While semilocal DFT can provide important information and has helped in first-stage screenings,\cite{broberg2023high} it remains essential to always verify these crude methods with a hybrid functional at the very least for the most promising materials. Indeed, similar to how semilocal DFT underestimates band gap, it tends to overpredict dopability, and its use without safeguard has led to false predictions of p-type doping. One pertinent example from the literature is the case of \ce{Cs4CdSb2Cl12}, one of a new class of quadruple perovskites initially predicted in 2018 as p-type TCs using DFT defect calculations without hybrid confirmation.\cite{xu2018prediction} However, when \ce{Cs4CdSb2Cl12} was synthesized the following year, p-type doping was not achieved; these authors fact-checked theoretical dopability using a hybrid HSE functional, and results suggested these materials were not actually highly p-type, with doping limited by Cd interstitials (see schematic defect formation energy diagrams in \autoref{fig:examples}).\cite{hu2020p} In response, the authors of the original study then performed follow-up hybrid calculations and revisited their original claim: \ce{Cs4CdSb2Cl12} could indeed be highly p-type, but only within a specific region of chemical potential space (Cl-rich and Cd-poor, with fast quenching during growth).\cite{xu2021exotic} To our knowledge, doping within this region still awaits experimental confirmation.

The field of high-throughput defects is moving fast with interesting developments in the use of methods compromising between DFT and full hybrid such as single-shot hybrid, as well as work on the more exhaustive search of the defect energy landscape.\cite{xiong2023high, thomas2024substitutional, mosquera2023identifying} In all cases, defect computations remain expensive and proxies have been highly sought out. Some recent studies have applied practical single-defect-based descriptors early in the screening, such as the hydrogen interstitial defect descriptor ("FEH")\cite{yim2018computational} or the oxygen-vacancy formation energy descriptor.\cite{kumagai2023computational} Other studies have used branch point energy (BPE, or $E_\mathrm{BP}$) as a descriptor,\cite{sarmadian2016easily, shi2017highthroughput} but we caution against using BPE as its predicting power is questionable.\cite{woods2018assessing}

% \begin{figure*}
%     \centering
%     \includegraphics[width=180mm]{comp-v-exp.png}
%     \caption{A comparison of computed vs. experimental properties for a representative set of p-type TCs in which optical and electrical properties have been reported. (a) Computed band gap compared to the range of reported experimental band gaps. (b) Computed hole effective mass $m^*_\mathrm{h}$ compared to maximum experimentally achieved gap. Data has been sourced from Woods-Robinson et al. and recent literature.\cite{woods2018assessing}}
%     \label{fig:comp-v-exp}
% \end{figure*}

\subsubsection{Thin films properties can diverge from bulk properties}

%%% mostly focused on the role of imperfect crystallinity in properties of thin films

% For the large majority of applications, a transparent conductor is required in the form of a thin coating. Although TC technologies based on nanowire networks or atomically-thin materials exist or have been proposed,
Practical TCs are normally required in the form of thin films with thicknesses on the order of a few hundred nanometers. Even if bulk properties are appropriately modeled, and if thin films can be grown, this requirement generates another possible source of disconnects between calculated and experimental properties (stages 4--5 in \autoref{fig:framework}). Most first-principles materials modeling methods assume a perfectly crystalline and infinitely periodic material. However, both assumptions often break down in thin film TCs, where non-equilibrium effects (e.g., non-equilibrium defects and solubility, strain and surface tension, substrate effects, influence from plasma, etc.) and nanoscale effects (e.g., grain boundary scattering, amorphous effects, quantum confinement) can dominate.\cite{yang2022big} 

Depending on the substrate and on the growth process, thin films can be epitaxial single-crystalline, epitaxial polycrystalline, non-epitaxial polycrystalline, and amorphous. Moving down this list, an increasing deviation from the properties of bulk single crystals is typically observed. The current generation of n-type TCs is either non-epitaxial polycrystalline (ITO, AZO, FTO) or amorphous (IGZO, ZTO, IZO).\cite{morales2017transparent} Therefore, it is unlikely that epitaxial single crystalline films can practically be used as p-type TCs, both because of the much higher cost and the requirement of a suitable substrate, which is usually dictated by the application (see section "Scalability and sustainability"). % Therefore, it is reasonable to expect experimental properties to deviate from theoretical, and this should be accounted for in modeling

The TC-relevant property that is most often affected by imperfect crystallinity is carrier mobility. At a hole concentration of \SI{e18}{cm^{-3}}, single-crystalline silicon has a hole mobility of about \SI{200}{cm^2V^{-1}s^{-1}}, whereas mobility in polycrystalline Si with \SI{120}{nm} and \SI{23}{nm} grains are about \SI{10}{cm^2V^{-1}s^{-1}} and \SI{1}{cm^2V^{-1}s^{-1}}, respectively.\cite{joshi1984mobility} At an electron concentration of \SI{e19}{cm^{-3}}, typical electron mobilities of ZnO are \SI{70}{cm^2V^{-1}s^{-1}} for single crystals, \SI{40}{cm^2V^{-1}s^{-1}} for epitaxial films, and below \SI{1}{cm^2V^{-1}s^{-1}} for non-epitaxial polycrystalline films.\cite{ellmer2010handbook} While the relative importance of grain boundary scattering decreases with increasing carrier concentration, grain boundary scattering can still remain a mobility-limiting mechanism even in the \SI{e21}{cm^{-3}} carrier concentration range,\cite{ellmer2010handbook} so it is relevant for TCs. The lower mobility of non-epitaxial polycrystalline films with respect to their epitaxial counterparts can also be due to a higher density of both point defects and extended defects. These considerations do not invalidate the screening approach, with the understanding that the computationally-determined mobilities are generally upper bounds for the mobilities achievable by polycrystalline thin-films. Large crystalline bulk mobility is a necessary but not sufficient requirement as grain boundary scattering could be an issue.\cite{marom2006contribution, ellmer2008carrier} Although grain boundary scattering is difficult to model from first principles, various simplified high-throughput approaches have been proposed and applied,\cite{ganose2021efficient, wang2022band, pohls2021experimental} and including it in screenings could be an interesting future avenue of research.

\begin{figure*}
    \centering
    \includegraphics[width=\textwidth]{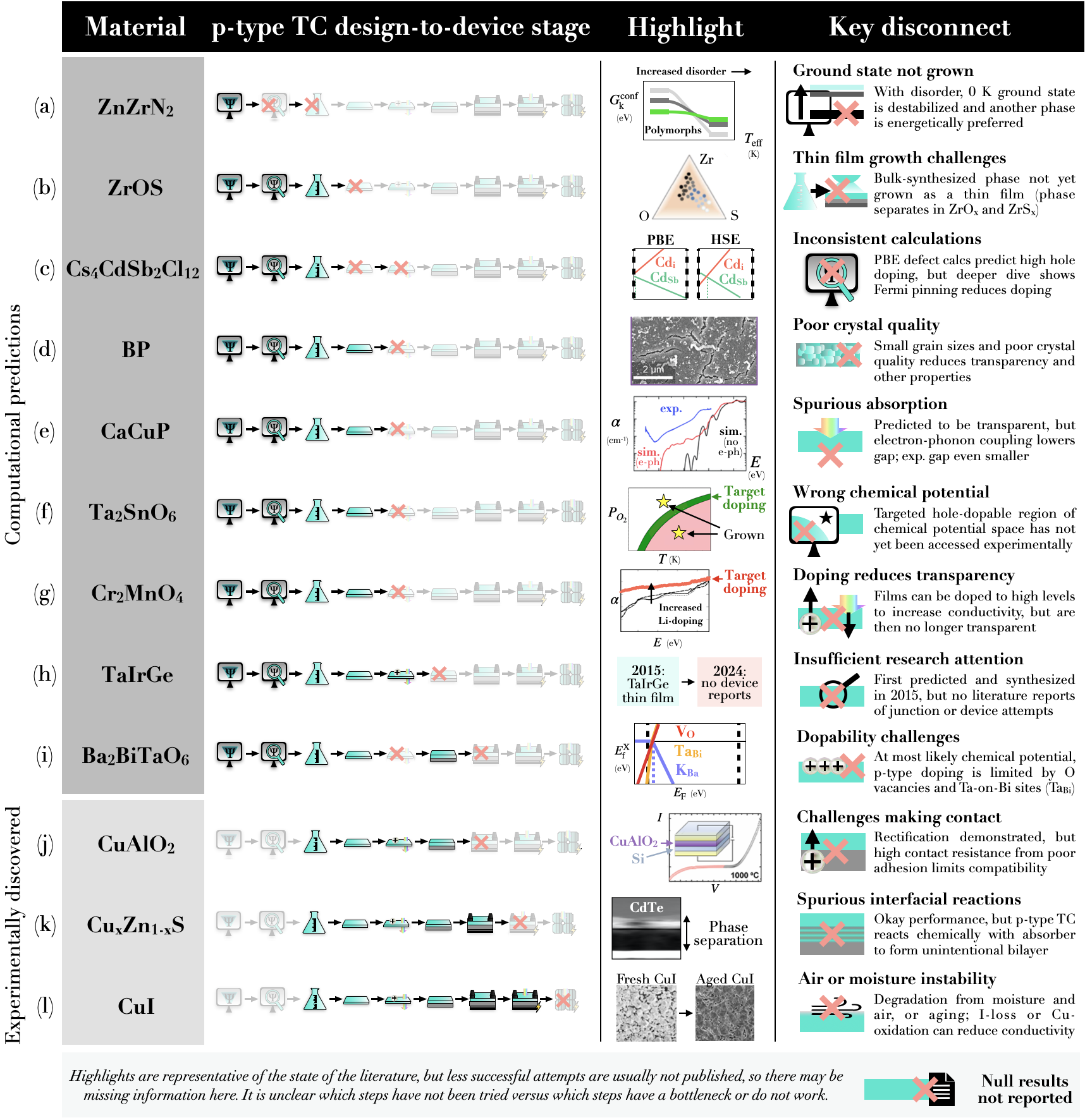}
    \caption{Examples of proposed and realized p-type TC candidates across the stages of their development process (as outlined in \autoref{fig:framework}). Materials (a--i) have been predicted computationally, while materials (j--l) were discovered first by synthesis and are included here as references. The materials are included as follows with the "highlight" column contains key results adapted from figures from the literature:
    (a) \ce{ZnZrN2} (Woods-Robinson et al.\cite{woods2022role}), (b) \ce{ZrOS} (adapted with permission of Monica Morales Masis, from Fioretti and Morales Masis\cite{fioretti2020bridging}, copyright 2020, SPIE), (c) \ce{Cs4CdSb2Cl12} (Hu et al.\cite{hu2019first}), (d) \ce{BP} (Crovetto et al.\cite{crovetto2022boron}),  (e) \ce{CaCuP} (Willis et al.\cite{willis2022prediction}), (f) \ce{Ta2SnO6} (adapted with permission from Barone et al.\cite{barone2022growth} Copyright 2022, American Chemical Society), (g) \ce{Cr2MnO4} (Nagaraja et al.\cite{nagaraja2014experimental}), (h) \ce{TaIrGe}, (i) \ce{Ba2BiTaO6} (adapted with permission of Geoffroy Hautier, from Dahliah et al.\cite{dahliah2020defect}, copyright 2020, Royal Society of Chemistry), (j) \ce{CuAlO2} (Ling, et al.\cite{ling2010color}), (k) \ce{Cu_{x}Zn_{1-x}S} (Woods-Robinson, et al.\cite{woods2020sputtered}), and (l) \ce{CuI} (Raj, et al..\cite{raj2019introduction}). Permissions are conveyed through Copyright Clearance Center, Inc., and all other figures have been adapted from a license with reuse permissions. Details about highlight figures and corresponding data tables are included in the Supplemental Material.\cite{si}}
    \label{fig:examples}
\end{figure*}

Another property that can be affected by imperfect crystallinity is optical transparency, especially when the fundamental band gap of the material is indirect and located at visible (rather than UV) wavelengths. In these cases, applicability of the material as a TC relies on the weakness of indirect optical transitions, resulting in negligible absorption in a thin-film sample because transitions involve both a photon and a phonon. Deviations from perfect crystallinity due to, e.g., closely spaced grain boundaries, extended defects, and inhomogeneity, are expected to relax the requirement for phonon participation in these transitions and, hence, increase the material's absorption coefficient. The increase in the absorption coefficient of silicon between its indirect and direct band gap with decreasing crystalline quality is a well-known example.\cite{jellison1993optical} A similar effect has recently been observed in polycrystalline BP and CaCuP films, predicted p-type TCs with direct gaps much larger than indirect gaps (see absorption spectra of \ce{CaCuP} in Figure 5).\cite{willis2022prediction, crovetto2022boron} In both cases, the experimentally reported absorption coefficient in nanocrystalline films is much higher than the calculated absorption coefficient using hybrid functionals and electron-phonon coupling, rendering them non-transparent, and an example of poor crystal quality in \ce{BP} is depicted in \autoref{fig:examples}. 

Due to the requirement of inexpensive, large-area thin-film samples for TC in PV applications, such non-idealities might be unavoidable in practice. One may argue that that some detrimental effects of polycrystallinity may vanish for large crystal grain sizes in the $\mu$m range, which can often be achieved in thin film samples, including In-based TCOs.\cite{ellmer2008carrier} For some TCs such as ZnO, FTO and large-grained CuI, optical scattering effects typically become more prominent with increasing grain size due to an increase in surface roughness. This is visually manifested as a "haze" effect, which can be beneficial in thin film PV for light management, but could also result in low transmittance.\cite{yamada2016truly} For reference, the grain size of commercial ITO films (n-type TCs) is often below \SI{30}{nm}.\cite{fanni2014c, kralj2023impact}

One overarching disconnect between bulk and thin-film materials follows from the fact that many thin-film growth techniques operate under strongly non-equilibrium conditions, where thermodynamics-based modeling methods may not be appropriate. Some examples are plasma-based techniques such as sputter deposition, pulsed laser deposition, and plasma-assisted chemical vapor deposition. Interestingly, these are some of the most commonly used deposition techniques for TCs, partially because the presence of energetic species in the plasma can promote higher dopant solubility at low temperatures. In these types of processes where both ions and neutral species are present with a range of energies, the atomic chemical potentials used in the calculations of defect formation energies are not easily translated to experimental process conditions. Thus, the dopability of a TC may be significantly different than expected. The mobility may also be affected, since it is often limited by defect scattering in highly doped materials.

\subsubsection{Challenges optimizing thin film growth and defects}

%%% Focuses on doping, chemical potential

Early synthesis attempts usually result in low-quality properties, which can be optimized and tuned with increased research efforts. For example, in the first reports of thin film n-type ITO in the 1970s conductivity was approximately 300 S cm\textsuperscript{-1},\cite{mehta1972sputtered} but after two decades of intensive research conductivity was optimized to nearly 2 orders of magnitude higher (with a record of $\sim$22,000 S cm\textsuperscript{-1}\cite{rauf1996structure}). Since most of the predictions of p-type TCs have been within the past decade and a given prediction has received far less attention than ITO, it is reasonable that development of predicted TCs that have been grown as thin films is stalled between stage 4 and 5 in \autoref{fig:framework}.

Growing a high performance p-type TC requires optimal control of growth conditions and thus chemical potentials. Many of the computationally identified candidates offer p-type doping only within a specific window of chemical potential. For instance, a well-adjusted oxygen chemical potential can be essential during growth, since highly-reducing conditions can create oxygen vacancies that compensate any holes. We note that this problem can be present even if an shallow acceptor is incorporated in the film. \ce{Ba2BiTaO6}, for instance, shows limited p-type doping because of the presence of oxygen vacancies (\ce{V_O}) and Ta-on-Bi anti-sites (\ce{Ta_{Bi}}), as shown in the defect formation energy diagram of \ce{Ba2BiTaO6} in \autoref{fig:examples} (adapted from Dahliah et al.).\cite{dahliah2020defect}

While computational screening uses some level of dopability, screening leads to focus only on materials that could potentially be p-type doped (i.e., that could exist in chemical potentials that are favorable for hole doping). There is not an easy relation between chemical potential and experimental growth conditions. While qualitative statements can be made --- i.e., lower oxygen chemical potential will be obtained by working in reducing atmosphere --- computations cannot tell what exact conditions (gas flow, temperature, etc.) to use experimentally. The computed chemical potential can be influenced by effects not taken into account (alloying, unknown phases or temperature effects). Similarly, source material impurities should be carefully checked; e.g., C and Si are dominant impurities in boron sputter targets, but are some of the most effective dopants in BP (Si impurities alone can lead to carrier concentrations \SI{e19}{cm^{-3}}.\cite{crovetto2022boron}) 

Any single growth technique will have limits in terms of attainable chemical potentials. A recent case in point is the growth of \ce{Ta2SnO6},\cite{barone2022growth} which has been computationally identified and reported as a potential p-type TCO (albeit with a small gap of 2.4 eV) and is highlighted in \autoref{fig:examples}.\cite{hu2019first, robertson2021doping} High quality single crystals grown by molecular beam epitaxy (MBE) did not yield any measurable conductivity, even after extrinsic doping attempts using K and Ti. The most likely reason for these failed doping attempts lie in the limits of oxygen chemical potential achievable in MBE. Oxygen vacancies act as hole-killers and need to be avoided under oxidizing conditions. However, the most oxidizing conditions available with this MBE process are still not oxidizing enough to enable p-type doping. It remains to be seen if other growth techniques will lead to p-type doping in \ce{Ta2SnO6}, and similar effects likely occur in other p-type TC candidates.

For potential TCs belonging to non-oxide families, a common experimental challenge (besides air stability, discussed earlier) is to ensure low levels of oxygen contamination in the films. This issue is particularly difficult to address when the target TC material contains highly oxophilic metals, such as the ones from Groups 1, 2, 3, 4, and Al. To address this problem, we recommend including "oxygen tolerance" calculations of promising TC materials, i.e., the calculated formation energy of oxygen substitutional and interstitial defects as a function of Fermi level. If any of these defects is found to be a donor with low formation energy, oxygen contamination is likely to limit the p-type dopability of the material. The development of more general oxygen-tolerance principles would also be welcome. For example, one may expect that oxygen substitution on the anion site is a donor defect in pnictides (one more valence electron) and an acceptor in halides (one less valence electron). If this is correct, oxygen incorporation would be detrimental for p-type conductivity in pnictides but it might be benign or even beneficial in halides. Beneficial effects of oxygen incorporation have been shown for CuI.\cite{storm2021evidence,crovetto2020water} Furthermore, the formation energy of these substitutional defects might generally be lower in materials containing an anion with a similar ionic radius to O$^{2-}$. Thus, one may intuitively expect nitride p-type TCs to be most negatively affected by oxygen contamination.\cite{talley2019synthesis}

As eluded to with these examples, optimization of TCs often involves introducing defects, disorder and off-stoichiometry in a tunable manner. Sometimes intrinsic doping is possible, but usually extrinsic dopants are needed to increase dopability. While most screenings include some consideration of dopability, often only intrinsic defects are considered, and detection of problematic hole-killers is targeted without providing the exact shallow dopant to use. However, in some cases dopants have been proposed, e.g., for \ce{La2SeO2} (Na)\cite{sarmadian2016easily} \ce{BP} (Mg, Si, Be, C)\cite{varley2017highthroughput},  \ce{Ba2BiTaO6} (K),\cite{dahliah2020defect} and \ce{CsCuO} (Na, K, Rb),\cite{kumagai2023computational}. Our experience is that defect compensation is often more limiting than the discovery of a shallow acceptor. There are a few rare cases where, even if a material does not have strong compensation, issues still arise in discovering the ideal dopant (e.g., n-type \ce{La2Sn2O7}\cite{hensling2021epitaxial}).

Moreover, defects and other impurities can influence absorption. Doping to degenerate levels can induce free-carrier absorption in TCs, reducing transparency at for long-wavelength photons. Often computational predictions target perfect stoichiometric materials, but promising undoped p-type TC candidates may result in trade-offs in properties upon doping, resulting in false positives. For example, an early screening study predicted \ce{Cr2MnO4} as a promising candidate, and selected Li as a p-type dopant.\cite{peng2013li} Synthesis of Li-doped \ce{Cr2MnO4} thin films indeed yielded p-type conductivities (though only up to 0.035 S cm\textsuperscript{-1}), but at high doping transparency is significantly reduced due to absorption of photons with energies below the band gap, as shown in \autoref{fig:examples}.\cite{nagaraja2014experimental} Recent computational studies have proposed computational methods to estimate plasma energy as a function of hole concentration,\cite{chen2022design} however methods to assess the impact of doping on transparency have yet to be implemented in high-throughput screenings.

Likewise, although screenings usually identify ordered compounds, the best known experimental TCs are off-stoichiometric solid solutions rather than dilutely-doped compounds. For example, ITO is approximately 5--10 at. \% tin, exceeding the dilute limit,\cite{frank1982electrical} and Al content in Al:ZnO (AZO) is approximately 2\%.\cite{crovetto2016performance} One workaround is to screen for cases in which a solid solution of the two compounds is amenable to high transparency. We have proposed this framework previously, and incorporated an "alloys database" tool to do so in the Materials Project.\cite{woods2022method} Additionally, in some cases TC properties could emerge non-linearly from tunability. For example, in p-type \ce{NiCo2O4} spinels \cite{chen2015p} disorder has been shown to reduce the hopping barrier and actually lead to increased conductivity compared to that of the ordered structure.\cite{exarhos2007cation} Finding methods to appropriately model and screen for dopants and tunability early on (stage 1-2) rather than during experimental optimization (stage 5) could lead to p-type TC breakthroughs, but this has not yet been done in practice.

\subsection{Novelty bias and confirmation bias}

Conventional n-type TCs such as \ce{ZnO}, ITO, etc., have had their growth techniques and properties fine-tuned over decades by international research teams and industries for numerous applications. Thus, it is unsurprising that the first results of a new predicted material may not have properties comparable to predictions. But, although it is unfair to compare a preliminary result with the a highly optimized result of n-type counterparts, this is still often done, and may lead to discarding a promising material that just requires optimization to show its real potential. Even if the target material is not synthesized in a particular study, this does not preclude the possibility of synthesizing these compounds using other techniques. Yet null results are usually not reported. This limits the ability for followup research to learn lessons from less successful results and for results to be reproduced to increase statistical significance, while increasing the chance of repeating mistakes and wasting resources. We have highlighted a few key exceptions; for example, Barone et al. reported being unable to synthesize conducting phases of \ce{Ta2SnO6},\cite{barone2022growth} but this report inspired follow-up calculations to suggest their MBE growth did not sample optimal chemical potentials.\cite{hu2023amorphous} To remedy this, amorphous \ce{Ta2SnO6} has been proposed,\cite{hu2023amorphous} and these studies inspire future work to explore different regions of phase space.

The initial synthesis of novel computationally-identified thin-film materials often entails sub-optimal growth conditions, such as chemical potential. This is particularly true for so-called "high-risk" novel materials, in which customized tools likely do not yet exist, necessitating makeshift adaptations of existing experimental tools. Often, growth chambers designed and previously (or concurrently) used for conventional materials like oxides are repurposed for the synthesis of multiple new materials. Such adaptations raise concerns about contamination, especially by volatile species like alkali metals, zinc, sulfur, or halogens, which can compromise the material's synthesizability, quality and defect sensitivity. For example, growing p-type phosphide TCs in chambers also used for sulfides may adversely affect hole concentration. "Exploratory" tools such as combinatorial sputter chambers can offer a wide range of elements and tunability,\cite{mittmann2024phosphosulfide, mcginn2019thin, talley2018implications, parzyck2023atomically, schnepf2021reactive} so they are particularly advantageous at stage 4 in Fig.~\ref{fig:framework}. However, exploratory thin film growth chambers accommodating volatile species are not particularly widespread, which is a barrier to enabling synthesis of new predicted p-type TCs. Often when predictions are made, one of the limiting factors is finding a synthesis team willing to risk attempting a novel compound with possibly exotic elements. In spite of these challenges, new high-throughput materials synthesis methods are on the rise, and could enable faster progression from a predicted TC (stage 1) to an optimized thin film (stage 5).\cite{tabor2018accelerating, moradi2022high, szymanski2023autonomous}

%%%%%%%%%%%%%%%%%
%% Lab to Cell %%
%%%%%%%%%%%%%%%%%

\section{Lab-to-device disconnects}

If a predicted p-type TC can be optimized to high performance in the lab, incorporating this material into an actual device will bring a set of new disconnects, some of which are summarized and schematically depicted on the right of \autoref{fig:disconnects}. Due to intrinsic disconnects from stages 1--5 and new interface challenges, so far very few predicted p-type TCs have been successfully reported as junctions and in solar cells (stages 6--9 in \autoref{fig:framework}). For example, although \ce{TaIrGe} was predicted and synthesized in 2015 as a thin film with high transparency and hole conductivity,\cite{yan2015design} nearly a decade later we are unable to find any reports of TaIrGe-based junctions or PV devices, suggesting either insufficient research attention or device attempts that have not been published (see \autoref{fig:examples}).

Not only are device reports of predicted TCs rare, but we were unable to find reports of \textit{any} p-type TC included as a layer for lateral hole transport, i.e., "electrode" in (a--c) from \autoref{fig:device-designs}, even as a small-scale proof-of-concept cell (stage 7). Although p-type \ce{Ba2BiTaO6} has been demonstrated as a p-n junction diode with n-type \ce{SrTiO3} (see \autoref{fig:examples}), \cite{shi2022modulation} demonstrating carrier transport and device potential, to our knowledge it has not been grown in a PV device. The only exceptions are \ce{NiO_x}, which is not technically a p-type TC due to low hole conductivity (although a recent report suggested external dopants could increase higher hole transport\cite{fan2022electronic}), and \ce{Cu_xS}, which is not very transparent, and in both cases devices have very low efficiencies.\cite{jacobsson2022open} From our understanding, p-type TCs have only been reported in solar cells as out-of-plane transport layers stacked next to n-type TC electrode or as continuous metal layers, as in the HTL of \autoref{fig:device-designs}(d), or as a semi-transparent bottom contact as in \autoref{fig:device-designs}(e). 

With these limitations in mind, in this section we reflect on some lessons learned from reported attempts to incorporate p-type TCs into PV-relevant devices as HTLs, generally following the device configuration shown in \autoref{fig:device-designs}(e). We use insights from several p-type TCs which were first discovered experimentally such as \ce{Cu_{x}Zn_{1-x}S}, \ce{CuI}, and delafossites \ce{Cu$M$O2}, as well as common non-TC hole transport materials (HTMs) such as \ce{NiO_x} and common n-type TCs (see \autoref{fig:examples}). Our expectation is that insights will be transferable to computationally predicted TCs once they can be optimized, and that when advances in material properties are achieved (stages 1--5), understanding these insights will enable more rapid scale-up into PV devices (stage 6--9).

\begin{figure*}
    \centering
    \includegraphics[width=\textwidth]{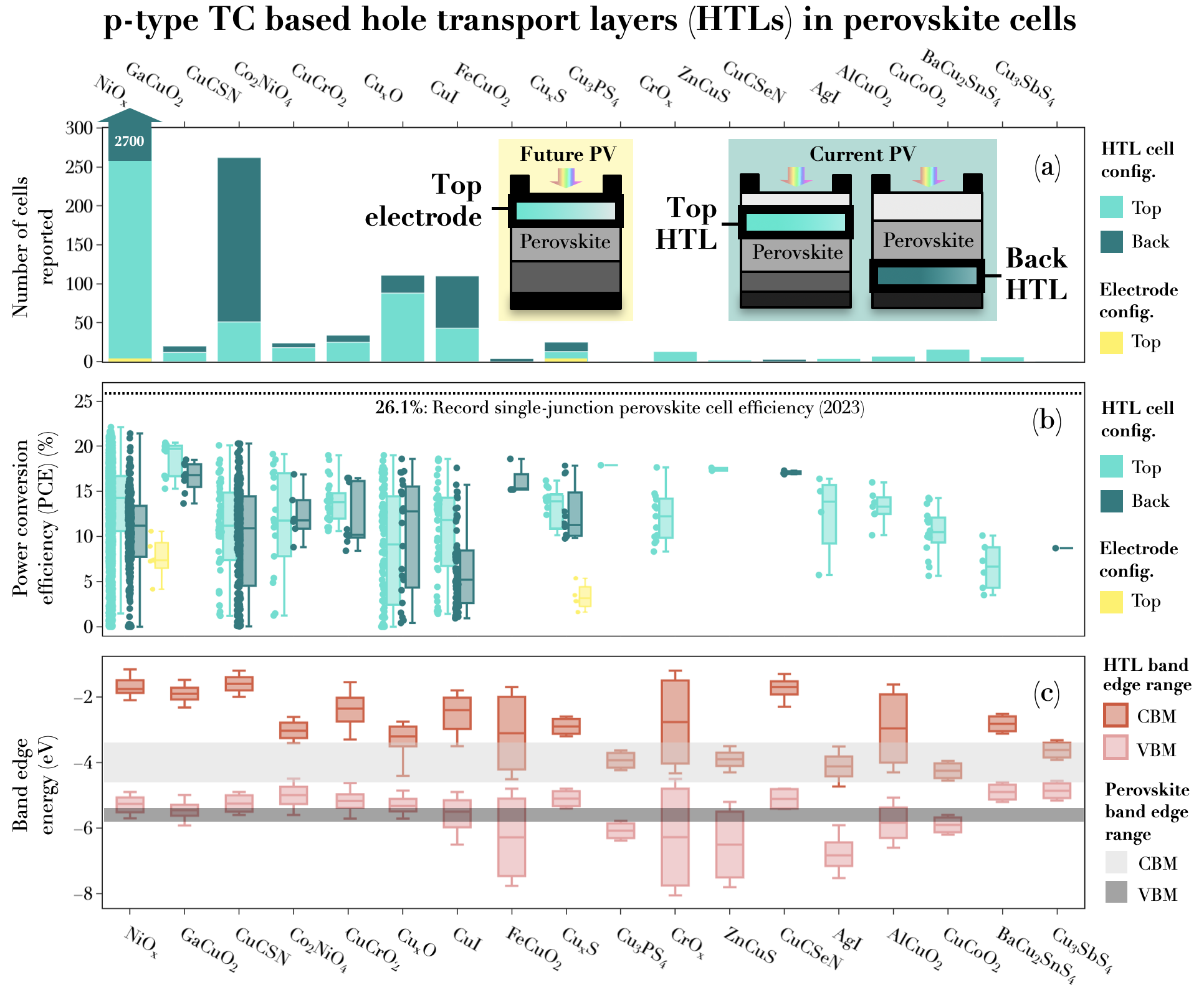}
    \caption{Summary of various p-type TCs used as hole transport layers (HTLs) in laboratory-scale single-junction perovskite solar cells, and (a) a histogram of the number of cells reported, (b) box plot of reported power conversion efficiency (PCE), and (c) the range of reported band edges from the literature. In (a) and (b), color refers to the cell configuration and whether the HTL is at the top or bottom of the cell (see \autoref{fig:device-designs} for details). Data to make this figure has been processed from the Perovskite Database\cite{jacobsson2022open} (containing studies from 2021 and earlier) and various literature reports, and is available as a table and \texttt{jupyter} notebook in the Supplemental Material.\cite{si}}
    \label{fig:perov-HTMs}
\end{figure*}

To illustrate the variety of p-type TC contact materials that have been recently incorporated into PV devices, and the magnitude of research efforts as well as what is currently lacking, \autoref{fig:perov-HTMs}(a) depicts a set of inorganic p-type TCs from the experimental literature which have been implemented as electrodes or HTLs in perovskite solar cell devices (\ce{NiO_x} is included as a benchmark). Data for this plot has been processed from the open access Perovskite Database.\cite{jacobsson2022open} For each reported cell, shading indicates whether it is are part of a double-layer top contact HTL stack, a double-layer back contact HTL stack, or a single-layer top electrode; notably, only two binary materials have been reported as a top electrode, and therefore this is an area for future research. Perovskite cells have been selected here as representative since the search for HTLs is an active area of research and therefore many p-type TCs have recently been attempted; we acknowledge the high hole mobility requirement of p-type TCs is not usually needed for HTLs (and some do not need very high transparency), so this test case represents a specific subset of TC design criteria.

\autoref{fig:perov-HTMs}(b) plots the distribution or reported power conversion efficiency (PCE) across cells, and materials are sorted by the highest reported PCE for a given material. The highest PCE is from delafossites \ce{Cu$M$O2} ($M$=Ga, Cr, Fe), spinel \ce{NiCo2O4}, mixed anion compounds \ce{CuCSN} and \ce{Cu3PS4}, and binaries \ce{Cu_{x}O} and \ce{CuI} (with various doping schemes). We note that none of these p-type TC materials were first predicted computationally; to our knowledge to date no predicted p-type TC has been used as a perovskite HTL. Additionally, these are results from lab-scale cells, not modules; none of the reported materials (except for \ce{NiO_x}) have been reported in a large-area module.

\subsection{Challenges at interfaces and junctions}

% Chemical / thermodynamic alignment

Additional complexity emerges at device interfaces, where a TC is in contact with other material layers, which can lead to serious optimization challenges. Before incorporating into a full device, sometimes new contacts are tested on a heterojunction or a "half-cell" to assess the interface and demonstrate rectification properties (stage 6). One of the first challenges to emerge at the interface of a p-type TC and its adjacent layer(s) is interfacial stability. The interface must be stable without reacting, decomposing or segregating into unintentional secondary materials or inducing trap states. For example, although \ce{Cu_{x}Zn_{1-x}S} is stable as a thin film, growing it as a back contact to CdTe solar cells induces complete decomposition of the \ce{Cu_{x}Zn_{1-x}S} layer into Cd-Zn-S and CuTe see phase separation from electron microscopy in \autoref{fig:examples}).\cite{woods2020sputtered} Although \ce{NiO_x} is commonly used as a HTM for perovskite solar cells, at certain processing conditions it has been shown to react with the perovskite layer and introduce an interfacial defect layer; similarly, \ce{NiO_x} in contact with Si in a SHJ cell likely introduces an interfacial layer of \ce{SiO_x}. The Materials Project's interfacial reaction calculator can be applied as a simple indicator of whether a reaction product is likely to form thermodynamically at an interface and, if so, the degree of instability (whether it will likely passivate or decompose). However even if an interface is thermodynamically stable, it can still degrade over time after operating in non-equilibrium conditions, such as in light-induced degradation (LID). For example, the morphology and doping of n-type top contacts can impact the degree of LID,\cite{hamelmann2016light} and, in perovskites, HTMs \ce{NiO_x} and PEDOT:PSS degrade upon light exposure,\cite{elnaggar2020decoupling} but this effect is nontrivial to predict without experiments.

% synthesis methods of adjacent layers are incompatible such that the interface degrades, or that - Some devices can tolerate imperfect interfaces, but this is difficult to predict. 

% Electronic alignment

When incorporating a novel p-type layer into a PV device, electronic band edges have to be aligned such that they enable desired junction properties. Ideally, the VBM of the p-type TC should be aligned to the VBM of the layer it extracts holes from. In addition, the CBM of the TC should be sufficiently shallow to form an electron-blocking barrier with the same layer. These criteria are important in particular for hole selective contacts for thin film PV, and for p-type layers in tandems. For example, in CdTe solar cells a rule-of-thumb is that the VBM offset of the p-type contact and CdTe absorber layer should be within 0.3 eV.\cite{liyanage2019role} \autoref{fig:perov-HTMs}(c) plots literature-reported band offsets for p-type TCs used as HTLs in perovskite solar cells, showing that for most TCs the VBM range is close to the range of various perovskite absorber layers (-5.42 eV to -5.77 eV, referenced to vacuum\cite{jafar2017fabrication}). However, precise band alignment may not be crucial if the TC is placed next to another highly-doped layer, or if it is a tunneling layer. In SHJ cells it has been shown that if the hole selective contact is not highly doped, its VB has to be aligned very close to that of its neighboring layer (within $\sim\pm$0.1 eV) to enable high efficiencies. On the contrary, if the p-type TC layer is highly doped above $10^{18}$ cm\textsuperscript{3}, high efficiency can be achieved within a wider tolerance to VB misalignment.\cite{woods2022role}

A combination of X-ray photoemission spectroscopy (XPS) and first principles slab calculations can be used to assess band offsets of new TCs. Uncertainties of DFT alignments are less than $\pm$0.3 eV with LDA and GGA-PBE (this is included as error bars in \autoref{fig:perov-HTMs}),\cite{dewaele2016error} and this error becomes problematic if a precise degree of alignment is required between layers.
Band offsets can differ depending on the surface morphology, including crystallographic plane, surface termination and defects, so there is often a mismatch between computed bulk band offsets and measured offsets in polycrystalline TCs.\cite{butler2019designing} Band offsets are usually not computed early in TC  screenings, but sometimes are assessed in computational deep dives (stage 2). Offsets for an individual material are not sufficient to simulate band bending when in contact with another material, so continuum simulations should be also performed to assess band bending and quasi-Fermi level splitting at a new interface.

% Physical/mechanical alignment

In addition to chemical stability and electronic alignment, fabrication of the p-type TC layer must form strong bonds and mechanically adhere to adjacent layers. Mechanical breakage at interfaces and delamination can induce detrimental voids or pinholes, which can result in increased electrical resistance and heat losses, and air or moisture gaps can facilitate degradation. These challenges are exacerbated in layered anisotropic p-type TCs such as delafossite \ce{CuAlO2}, as highlighted in \autoref{fig:examples} for \ce{CuAlO2}-Si heterojunction diode devices,\cite{ling2010color} where high contact resistance due to poor adhesion has been a major barrier to scaling solar devices. It is expected this may be the case for predicted layered structures such as \ce{[Cu2S2][Ba3Sc2O5]}.\cite{williamson2020computationally}. Some PV applications such as perovskites are moving towards flexible substrates, so designing p-type TCs layers to avoid mechanical breakages will become even more important as these devices come to market.

\subsection{Practical solar cell design}

% Defects at interfaces, pinning, passivation, other challenges

When a solar cell is fabricated using a new contact material (stage 7), cell performance is determined not only by intrinsic material properties but also by interfacial properties, in particular electronic band alignment, and the possible presence of interfacial defects. Surface passivation of contacts and absorbers is in fact an important line of research in thin film and heterojunction-based PV, however, this topic goes beyond the purpose of this review.\cite{salome2018passivation, aydin2019defect} Defects at interfaces can have different origins, from chemical intermixing to the presence of dangling bonds, among others. However, the role of these defects are rarely considered in early-stage materials design. Therefore, an open challenge is the development of screening metrics to assess interfacial or pinning defects that may arise at contact interfaces, and how they may limit performance. To assess this, it is important to also pre-define the type of absorber material and assess its properties simultaneously. Integrating methods that merge defect calculations and materials predictions with continuum interface simulations in solar cell contact layers could provide valuable insights into the behavior of device-relevant defects under actual operating conditions.\cite{jones2022modelling}

% Growth method, thermal budget

Besides engineering defects, another important consideration for device design is optimizing fabrication method, including designing for temperature stability (most devices have a "thermal budget"). A growth method that works well for synthesizing a single film on a substrate may not be appropriate for a full device stack. In thin film solar cells, contact layers with low thermal budgets of a few hundred degrees C are usually preferable; the process temperature of the TC should not be so high to cause unwanted reactions in the bulk of the absorber or at the interface. However, many of the synthesis recipes for the best reported p-type TCs higher temperatures during some step of the process.\cite{morales2017transparent} For example, Mg-doped \ce{CuCrO2} grown by sputtering is one of the highest performing p-type TCs but requires a synthesis temperature of 750 $\degree$C, which limits its practicality for devices.\cite{nagarajan2001p} However, recent developments have enabled low temperature \ce{CuCrO2} synthesis and incorporation at an HTL in perovskite cells with high efficiency (see \autoref{fig:perov-HTMs}).\cite{yang2019multifunctional} Furthermore, plasma-assisted techniques such as sputtering, pulsed laser deposition (PLD), and plasma-enhanced chemical vapor deposition (PECVD) are often used to deposit TCs, and these methods can be harsh on the underlying material; however, it has recently been shown that oxide TCs can be deposited on perovskites with very low damage by PLD.\cite{smirnov2021scalable, soltanpoor2023low} P-type TC films for PV applications that will operate in the sun also need to be stable under UV irradiation, both in their bulk form and at their interfaces, though this is less important in superstrate devices in which glass absorbs much of the UV. It is also important to consider the lifetime of the device when designing materials to be incorporated, for example whether migration of dopants or segregation into secondary phases will occur over time.

% multiparameter optimization

In PV contacts, the tolerance to non-ideal properties such as electronic misalignment depends strongly on device architecture. Tolerances of a TC property within a given device may depend on multiple material and junction properties and therefore may be better described by a multi-objective optimization scheme rather than a serial screening “funnel”. For example, precise band alignment may not be as relevant if the TC is placed next to another highly-doped layer, or if it is a tunneling layer. In SHJ cells it has been shown that if the hole selective contact is not highly doped, its VB has to be aligned very close to that of its neighboring layer (within $\sim\pm$0.1 eV) to enable high efficiencies. However, if the p-type TC layer is highly doped above $10^{18}$ cm\textsuperscript{3}, high efficiency can be achieved within a wider tolerance to VB misalignment.\cite{woods2022role} Other properties besides alignment also depend on the device configuration. Perhaps most trivially, the wavelengths to which a material must be transparent depends on the absorption spectrum of the absorber layer. Similarly, the high mobility requirement and threshold varies dramatically across absorber technologies and device designs.

% Competition with well-optimized materials

Similar to the novelty bias mentioned previously, one practical challenge when incorporating new materials into solar cell devices is that device architectures are optimized around conventional materials. For example, SHJ solar cells have been highly optimized using a hydrogenated boron-doped amorphous silicon (B-doped p-a-Si:H) and ITO bilayer as the hole-selective top contact, with thicknesses and dopings of these layers and adjacent layers tuned specifically to work well with these materials. Thus, one cannot simply remove a highly optimized material (e.g., a-Si), insert a new layer (e.g., novel p-type TC) into the same device stack, and expect it to deliver optimal properties, even if that new material has high transparency and conductivity. Rather, the entire device may have to be re-optimized to accommodate the new architecture, energetic alignment, and emergent defects from inserting this new layer. However, this can be tedious when working with a large quantity of candidates in high-throughput methods, and thus this is not typically done. As a result a feedback loop ensues: device layers are limited to a small set of materials (that is, small compared to the size of computational databases) that are well-optimized, heavily characterized, easy and cheap to grow, and therefore are more likely to work well in a given device architecture. Then, more resources are invested in further optimizing these already well-optimized materials, and new materials such as predicted p-type TCs have a larger barrier to entry.

\subsection{Scalability and sustainability}

% %  Scaling challenges

To our knowledge, a high-performance solar module using p-type TC contacts has not yet been fabricated. However, if a high-performance p-type TC can someday be successfully integrated into a small-scale laboratory cell, the next stage toward commercialization would involve scaling up to full-sized module devices (stage 9 in \autoref{fig:framework}), which would introduce a new set of challenges. Module performance tends to lag behind cell performance, namely due to the requirement for larger surface area deposition and lower tolerance for non-uniformities.\cite{lee2017review} Challenges such as large-area conformity, uniform crystallinity, and strong adhesion without pinholes are exacerbated for thinner films,\cite{abbott2023practical} and since contact layers tend to be thinner than absorber layers in solar cells there is likely less tolerance in contacts for such nonidealities. TC layers in modules also need to be stable for decades and maintain high transparency and electrical conductivity while minimizing resistance losses and avoiding localized performance drops due to defects.

%% Deposition and cost

Several deposition methods used regularly to synthesize TCs in lab-based devices are non-uniform by design and challenging to scale, such as spin coating and pulsed laser deposition (PLD), though wafer-scale PLD of ITO, Zr:\ce{In2O3}, and SnO2 for perovskites or SHJ cells has recently been demonstrated.\cite{smirnov2022wafer, zanoni2022ito, soltanpoor2023low} Other deposition methods such as sputtering are regularly adapted for large-area deposition, but require re-optimization and often result in performance lower than small cells. ITO is usually deposited by large-area sputtering in SHJ, perovskite, and SHJ-perovskite tandem modules,\cite{roffeis2022new} while in CdTe modules the n-type top contact FTO is usually deposited by CVD and can be done by the glass manufacturer \cite{scarpulla2023cdte}. Several studies have demonstrated large-area fabrication of buffers and contacts such as \ce{NiO_x} for perovskite module applications.\cite{li2017overcoming} However, scaling CuI while maintaining good performance remains an open challenge towards commercialization, largely due to instabilities from moisture, oxygen, and elevated temperatures\cite{liu2021engineering} (see degraded 6-month-aged sample in \autoref{fig:examples}\cite{raj2019introduction}), and similar challenges apply to other inorganic HTLs. Even if physically possible to scale, many high-vacuum synthesis techniques such as MBE are simply too expensive and labor-intensive to use commercially. Developing low-cost, module-scale synthesis techniques such as roll-to-roll printing early on in the p-type TC "design-to-device" development process could enable more rapid integration into next-generation modules. Incorporating economic modeling such as techno-economic analysis can help guide integration of new materials into low-cost devices.\cite{zafoschnig2020race}

Interdisciplinary collaborations between academia, national labs, and industry could help address scaling challenges. Researchers might assess the effects of large areas and conformity on the properties of early-stage TCs, whether low-cost, scalable deposition processes are suitable for a given TC, and how film quality trades off with cost and scalability when moving from epitaxial or single-crystal films to polycrystalline or amorphous films (realistic for large-scale processing). National labs such as NREL could set up testing and standardization protocols for large-scale TCs, and platforms to share data. Although thin film PV companies have likely explored emerging p-type TCs without publishing, successful public-private partnerships (e.g., First Solar’s p-type back contact research collaborations\cite{gorai2023search, perkins2023collaborative}) demonstrate the potential for joint efforts to advance module development. Continued collaboration could accelerate commercial scaling of TCs by linking the critical needs of industry with the capabilities and limitations of basic research.

%% Sustainability

Although the ultimate goal of scaling PV is to mitigate climate change, input and outputs from production and deployment may cause unintentional consequences such damage to ecosystems and human health, and there is an ethical responsibility to minimize harm as we scale new materials such as TCs.\cite{owen2013framework,wender2014anticipatory} Techniques such as life cycle assessment (LCA) can be used to identify key sources of negative impacts within a technology's life cycle and guide research directions to reduce harm.\cite{pennington2004life} PV has much lower environmental impacts than fossil fuel technologies due to negligible inputs and outputs during operation, but impacts do emerge from mining, refining, manufacturing, and decommissioning.\cite{nrel2021life} Mining of critical PV materials can lead to resource depletion, human health impacts, and ecotoxicity, and has motivated shifting towards “earth-abundant” materials,\cite{bergesen2014thin} such as moving away from indium in n-type TCs. However, in all PV devices most life cycle GHG emissions are from the energy intensity of material production rather than mining.\cite{iea2023energy} In conventional PV emissions are primarily from refining solar-grade silicon, but in emerging PV many of these manufacture-stage emissions are also from contact layers. For example, in perovskite tandems, fabrication of contacts and buffers such as Spiro-MeOTAD and sputtered ITO contribute significantly to greenhouse gas (GHG) emissions from the cell.\cite{roffeis2022new} Thin film PV tends to yield lower impacts compared to Si-based PV,\cite{fthenakis2009sustainability, rashedi2020life} which motivates scaling thin film PV with high-performing heterojunction contacts, so designing contact materials with low energy intensity processing could significantly reduce life cycle impacts.

Historically, LCAs are conducted after technologies reach commercial scale, which can lead to “technology lock-in” that makes it challenging to replace harmful materials.\cite{wender2014anticipatory} While there have been instances of course correction and phase-out --- for example, recent efforts to reduce Co-content in \ce{LiCoO2}-based batteries and the formation of NMC in \ce{LiNi_{$x$}Mn_{$x$}Co_{$y$}O2} to address cobalt supply chain hazards (which still rely on lithium and nickel)\cite{lee2022can} --- some of this lock-in is unavoidable if we want to rapidly deploy renewable energy. One such trade-off is that delaying deployment to wait for PV with lower embodied carbon would result in more emissions than installing existing PV, so it is important that conventional Si (e.g., PERC) is rapidly installed.\cite{husein2021delayed}

Low-TRL emerging PV technologies are years or even decades away from commercialization, but offer serious opportunities for sustainability improvements. These are technologies in which TCs play an important role, so designing new contacts for emerging PV with low environmental impacts could make a big impact. Assessing impacts early in the materials design phase is challenging,\cite{putsche2023framework} particularly when the synthesis route is unknown, and especially for p-type TCs since understanding their underlying physics and design criteria is still lacking. Computational materials discovery often assumes optimization can reduce impacts, e.g., from replacement of elements or reduction of processing energy. For example, a computationally-predicted p-type TC (\ce{BeSiP2}) has recently been proposed with a new design criteria, forbidden optical transitions; this material contains Be and is impractical to scale, but this design criteria may be applicable to an earth-abundant version that has not yet been discovered,\cite{woods2023designing} and being too restrictive could lead to missed opportunities. However, we should keep in mind that there is always a certain amount of risk with this approach since later-stage replacements can be difficult. For example, despite significant research efforts, replacing a toxic element in perovskite absorbers (Pb) with a non-toxic element (e.g., Sn, Bi) has not yet yielded high enough performance.\cite{ke2019prospects}

The following question arises: how can materials scientists integrate life cycle thinking throughout the full process of materials discovery and design to prevent "locking-in" technologies with detrimental impacts? This challenge calls for the development of early-stage metrics and strategies for embedding sustainability early in the materials design process, such as accounting for material criticality, embodied carbon, or social impacts from mining and refining in elemental precursors, or estimating possible energy consumption during manufacturing.\cite{putsche2023framework} Inspiration can be drawn from recent approaches like emerging materials risk analysis,\cite{horgan2023development} "upscaling" manufacturing assessments,\cite{weyand2023scheme} and the framework of anticipatory LCA,\cite{wender2014anticipatory,wender2014illustrating} as well as the development of open-source, open-access LCA infrastructure such as the \texttt{brightway} framework that can connect with materials discovery tools.\cite{mutel2017brightway} Moreover, we encourage the continued development of new metrics — assessing for embedded carbon, co-product assessment, manufacturing impacts, and end-of-life considerations — that can be incorporated into the design of p-type TCs for emerging PV and sustainable materials discovery in general.

%%%%%%%%%%%%%%%%%
%%% Takeaways %%%
%%%%%%%%%%%%%%%%%

\section{Insights for future design of p-type TCs}

This article has reviewed various disconnects throughout the materials discovery process of p-type TCs, spanning from computational design to material synthesis and from lab-scale devices to scalable sustainable modules. We have compiled and made available an extensive set of literature data tables to help researchers identify and explore such disconnects.\cite{si} To address these disconnects to design commercially-relevant p-type transparent conductors, we propose the following recommendations for materials designers:

\begin{enumerate}

    \item{\textbf{Improve and benchmark descriptors for p-type TCs, especially for band gap, dopability, mobility} Since the first high-throughput screening for p-type TCOs appeared a decade ago, first principles computational methods have improved tremendously and computing power has increased. More accurate assessment of band gap and mobility are emerging with different levels of accuracy and cost. Assessment of defects and dopability is still among the most expensive computations to perform, but they have been automatized and tractable high-throughput defect approaches are emerging. Predictive low-cost descriptors of dopability have not yet been developed, and including transport properties such as grain boundary scattering at early stages is not common, but both of these pursuits would accelerate screening tremendously. In the last decade available data from which to screen has also grown tremendously, such as in the Materials Project and other large databases, and opportunities to use machine learning to speed up the screening process are naturally emerging from this data.}
    
    \item{\textbf{Judiciously assess whether predictions are actually synthesizable.} Before lab-based synthesis, higher-accuracy metrics should be used to assess synthesizability of emerging screening candidates. To closer approximate synthesizability of realistic thin film and interface defects, computational screenings should seek better metrics than $E_\mathrm{hull}$ alone. The ability to move from thermodynamic metrics to assessment of synthesizability in screenings is an important future frontier of computational materials design.}
    
    \item{\textbf{Establish standards and consensus on how defect calculations are run and interpreted.} There are inconsistencies across the literature in terms of which corrections to use, and qualitative interpretation of formation energies and Fermi level pinning. To avoid inconsistencies and assist comparison across the literature, quantitative assessment of carrier concentration is recommended rather that simply stating whether a material is "dopable." Following lessons learned from Hu et al.'s quadruple perovskite study about the "importance of reasonable band gap description and chemical boundary determination in predicting defect thermodynamics,"\cite{hu2020p} is recommended to standardize running hybrid defect calculations to confirm p-type dopability.}

    \item{\textbf{Screen for tunability and tolerance ranges, rather than a single compound with a single property.} Tuning stoichiometry and doping in the lab can introduce effects not typically considered in bulk calculations, so techniques to simulate tunability would be useful. While very important in practical materials, the effects of alloying and off-stoichiometry are not often not taken into account early on in high-throughput studies, so addressing this gap could lead to new TCs and other material predictions. Automated experiments, though not yet used in p-type TC discovery as far as we know, are a promising future direction to incorporate tunability by linking feedback loops of experimental synthesis and characterization to machine learning and computational data. This could enable rapid exploration of chemical potential space and stoichiometry to limit false positives and identify nonlinearities.}

    \item{\textbf{Be judicious whether metrics and materials are device-appropriate.} Searches should be tailored to specific applications, rather than a generic "high-performance" p-type TC. Some properties matter less than others depending on application, and by using generic screenings useful targeted materials may be overlooked. For example, for some devices there may be trade-offs between properties such that achieving optimal performance is a multi-parameter optimization problem (e.g., carrier concentration and band alignment can be coupled). Assessing these trade-offs may require iterative research and learning between stages 1--9 in \autoref{fig:framework}, and synergizing high-throughput computation with multi-scale modeling and device testing.}

    \item{\textbf{Develop sustainability and scalability metrics to incorporate early in the materials design process.} To avoid "locking-in" technologies with detrimental environmental effects, materials scientists should integrate life cycle thinking throughout the design process, drawing inspiration from recent methodologies like emerging materials risk analysis and anticipatory life cycle assessment. Development of early-stage metrics assessing embedded carbon, co-product impacts, manufacturing impacts, and end-of-life aspects could help guide sustainable materials discovery in the field of emerging PV and beyond.}

    \item{\textbf{Conduct more interdisciplinary analyses to bring computed materials into devices.} Historically, separate research teams have performed computations, synthesis, fabrication, and scale-up modeling, which can lead to communication barriers. However, there is a recent trend towards connecting theorists, experimentalists, and analysts on a single project. Ultimately, more communication and collaboration between scientists across various stages of materials discovery and various stages of technology readiness can help address these disconnects and lead to more effective design.}

\end{enumerate}

The goal of materials discovery is to predict a new material from first principles that can ultimately scale up and be used in devices such as solar panels for more efficient and sustainable technological development. Although we have made major progress in the field of predicting p-type TCs and scaling into photovoltaic devices, there is still a large gap between the research stage of proposing a new candidate and actually incorporating the candidate into a device. In this perspective, we have highlighted some major disconnects from the past ten years of research into p-type TCs, as well as insights to guide new research into designing, synthesizing, and scaling materials. We hope that these insights, applied not only to future design of p-type TCs but towards materials design in general, can help catalyze more targeted and rapid discovery of materials to advance sustainable energy technologies.

\section*{Acknowledgments}

This work was supported by the U.S. Department of Energy, Office of Science, Office of Basic Energy Sciences, Materials Sciences and Engineering Division under Contract No. DE-AC02-05-CH11231 (Materials Project program KC23MP). R.W.R. was supported by the U.C. Berkeley Chancellor's Fellowship and the National Science Foundation (NSF) Graduate Research Fellowship under Grant No. DGE1106400 and DGE175814, and by University of Washington's Clean Energy Institute. G.H. acknowledges support from the U.S. Department of Energy, Office of Science, Basic Energy Sciences under award number DE-SC0023509. The work of A.C. was supported in part by a research grant (42140) from VILLUM FONDEN and co-funded by the European Union (ERC, IDOL, 101040153). Views and opinions expressed are however those of the authors only and do not necessarily reflect those of the European Union or the European Research Council. Neither the European Union nor the granting authority can be held responsible for them. 
% The views expressed in the article do not necessarily represent the views of the DOE or the U.S. Government.

\section{Author contributions}

We highlight the author contributions to this study using the CRediT taxonomy.

R.W.R.: Conceptualization, Methodology, Data Curation, Visualization, Investigation, Writing - Original Draft, Writing – Review \& Editing, Funding Acquisition, Project Administration;
M.M.M.: Methodology, Investigation, Writing – Review \& Editing, Funding Acquisition;
G.H.: Methodology, Investigation, Writing – Review \& Editing, Funding Acquisition;
A.C.: Methodology, Investigation, Writing – Review \& Editing, Funding Acquisition;

% \section*{Supplemental Information}

% A Supplemental Information file is included, alongside a corresponding supplemental table with the following data:

% \begin{itemize}
%     \item Computational predictions of p-type TCs.
%     \item Experimental reports of predicted p-type TCs.
%     \item Parsed literature review by material.
%     \item Benchmarking of computational vs. experimental reports.
%     \item Literature reports of valence and conduction band offsets in p-type TCs.
%     \item Reported performance, properties, and metadata of perovskite solar cells with p-type TCs used as contacts, parsed from the Perovskite Database.\cite{jacobsson2022open}
% \end{itemize}

% \noindent A series of \texttt{jupyter} notebooks to generate \autoref{fig:predictions}(c), \autoref{fig:perov-HTMs}, and SI figures is included. Selected supplemental data is also available via on the MPContribs platform.

% \bibliographystyle{unsrt}
% \bibliographystyle{apsrev4-2}
\bibliographystyle{ieeetr}
\bibliography{main.bib}

\end{document}

%%%%% TITLE %%%%%

\title{Supplemental Material for "From design to device: challenges and opportunities in computational discovery of p-type transparent conductors"}

\author{Rachel Woods-Robinson*\textsuperscript{1,2,3}, Monica Morales-Masis\textsuperscript{4}, Geoffroy Hautier\textsuperscript{5}, Andrea Crovetto\textsuperscript{6}}

\affiliation{\textsuperscript{1}Clean Energy Institute, University of Washington, Seattle, WA, 98195 United States, \textsuperscript{2}Materials Sciences Division, Lawrence Berkeley National Laboratory, Berkeley, CA, 94720 United States, \textsuperscript{3}Materials Science Center, National Renewable Energy Laboratory, Golden, CO, 80401 United States, \textsuperscript{4}MESA+ Institute for Nanotechnology, University of Twente, 7500 AE Enschede, the Netherlands, \textsuperscript{5}Thayer School of Engineering, Dartmouth College, 14 Engineering Dr, Hanover, NH, USA, \textsuperscript{6}Centre for Nano Fabrication and Characterization (DTU Nanolab), Technical University of Denmark, 2800 Kongens Lyngby, Denmark}

\date{\today}

\maketitle

\section{Computational predictions}

To accompany this manuscript, we have performed a literature review of p-type TCs which have been predicted using computational approaches, and selected results from synthesis attempts of predicted materials. A data table summarizing key findings and properties of these studies is provided as a supplemental spreadsheet.  This data set represents the studies we were able to find through a literature review, but likely is not comprehensive of every computationally studied p-type TC.

The first table contains 380 literature reports of computationally predicted and studied p-type TCs, which represent 261 unique predicted structures. The data from this table has also been released by the MPContribs framework and is available on the website (https://next-gen.materialsproject.org/contribs) or the MPContribs API, and anybody can add to it if interested.

\begin{figure*}
    \centering
    \includegraphics[width=\textwidth]{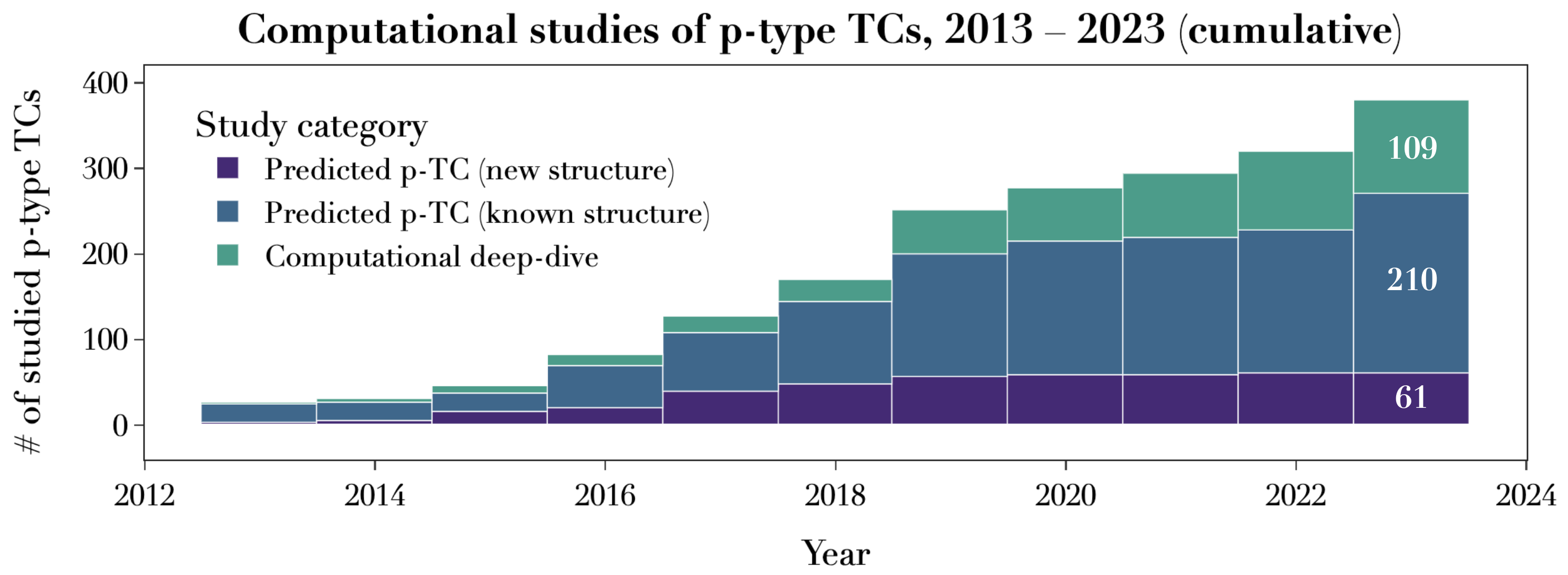}
    \caption{A cumulative timeline of published computational predictions of p-type TCs, from 2013 – 2023. Shading differentiates whether the study predicted a new structure for the first time (purple), whether it predicted a known crystal to be a p-type TC for the first time, or whether it  studies a previously predicted p-type TC using higher levels of theory than previous studies to assess optoelectronic properties.}
    \label{fig:pTCs-over-time}
\end{figure*}

In the manuscript, this data is used to make Figure 1a. Here in \autoref{fig:pTCs-over-time}, we report the number of studied p-type TCs as a function of time, from 2013 to 2023 (note that in several cases, a single literature report includes multiple predictions, and so these cases yield multiple "studied p-type TCs" in \autoref{fig:pTCs-over-time}). This shows that the number of predicted compounds has grown over time. The shading corresponds to whether a material from a given study is (a) has been predicted as a new crystal structure for the first time \textit{and} as a promising p-type TC candidate for the first time (purple shading), (b) is already a known crystal structure but is predicted as a promising p-type TC candidate for the first time (blue shading), or (c) has previously been predicted as a p-type TC candidate but is studied with higher degrees of theory to better estimate properties (green shading, "deep dive"). As of 2023, this data set consists of approximately 61 of (a), 210 of (b), and 109 of (c).

\section{Experimental reports}

\begin{figure*}
    \centering
    \includegraphics[width=180mm]{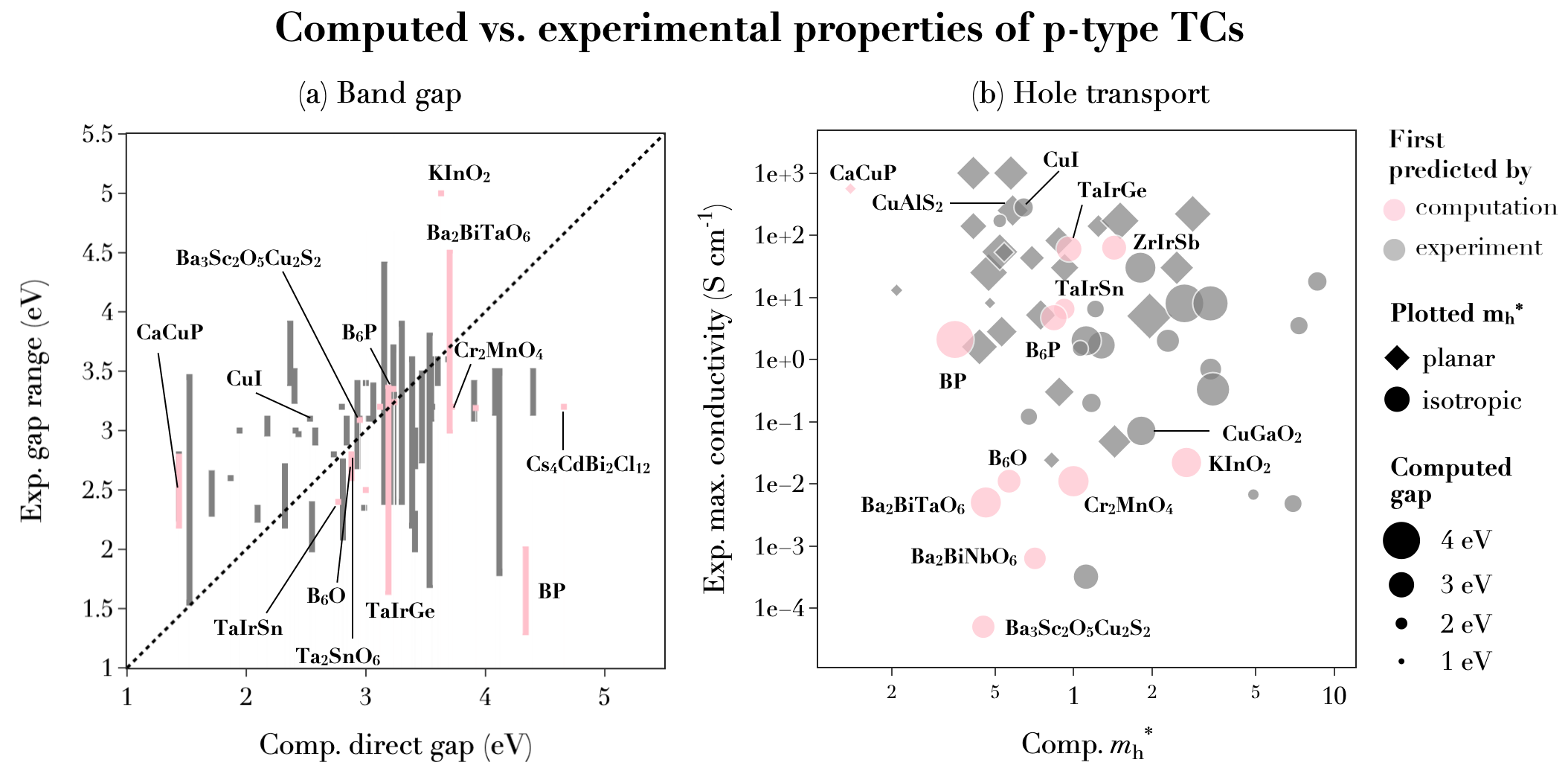}
    \caption{A comparison of computed vs. experimental properties for a representative set of p-type TCs in which optical and electrical properties have been reported. (a) Computed band gap compared to the range of reported experimental band gaps. (b) Computed hole effective mass $m^*_\mathrm{h}$ compared to maximum experimentally achieved gap. Data has been sourced from Woods-Robinson et al. and recent literature.\cite{woods2018assessing}}
    \label{fig:comp-v-exp}
\end{figure*}

To demonstrate how predicted properties compare to experimentally realized results, \autoref{fig:comp-v-exp} plots computed direct band gap versus the range of reported experimental gap in panel (a), and computed hole effective mass versus the highest reported conductivity in panel (b). The pink markers represent computationally predicted p-type TC candidates, many of which have been highlighted throughout the manuscript. The data from this plot is also derived from the supplementary table, with a variety of references cited. In several cases of predicted p-type TCs grown in the lab, properties falls short of the computationally predicted ideal. In others, experiment yields a wide range for a given property, in particular for reported band gaps such as \ce{TaIrGe}.

\section{Example materials}

\begin{figure*}
    \centering
    \includegraphics[width=150mm]{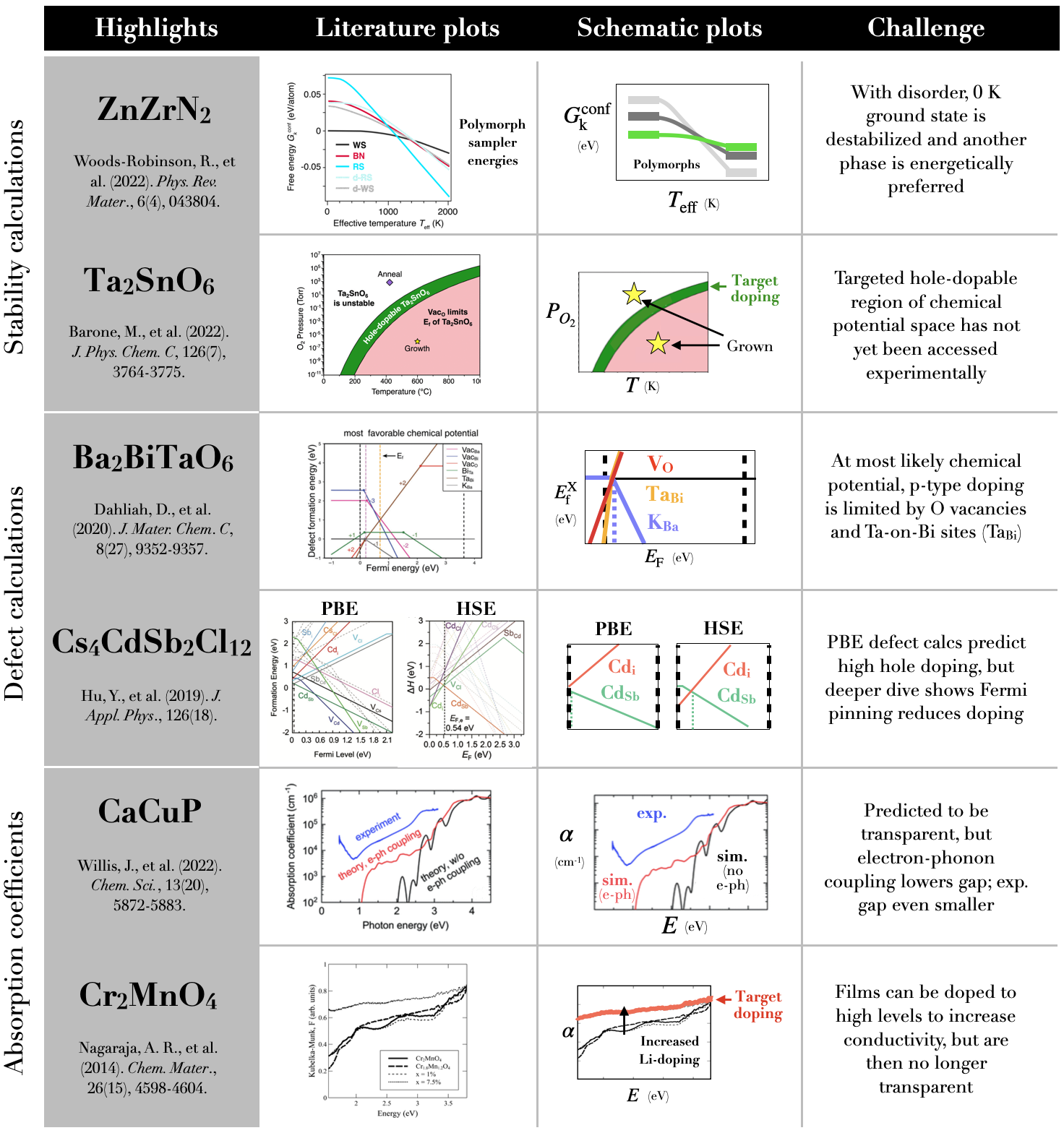}
    \caption{Example schematics for six materials from Figure 5 in the manuscript, with the plot from the literature depicted in the second column and our simplified schematic plot from the manuscript in the third column. Figures are reproduced and adapted with permissions as reported in the manuscript.}
    \label{fig:lit-v-schematic}
\end{figure*}

Here, we discuss details of highlighted experimental materials from Figure 5 in the manuscript, especially those pertaining to representative figures from the "Highlight" column. Plots corresponding to each schematic are reproduced from the literature in \autoref{fig:lit-v-schematic}. We highlight six of the examples of computationally-predicted p-type TCs to elaborate on specific methods: formation energy stability calculations (a,b), defect calculations (c,d), and absorption calculations and measurements (e,f). We note that materials highlighted are \textit{not} selected to claim they are the most promising p-type TCs, but rather they are selected as useful case studies to demonstrate a particular disconnect.

\subsection{Stability calculations}

\subsubsection{\ce{ZnZrN2}}

First, the plot for \ce{ZnZrN2} corresponds to free energy plots using the polymorph sampler method described by Stevanović et al.\cite{stevanovic2016sampling} The x-axis corresponds to the "effective temperature", which is a proxy for the degree of "cation disorder" within a crystal system, and the y-axis corresponds to the free energy. Each trace corresponds to a different polymorph in the \ce{ZnZrN2} stoichiometry: "wurtsalt" (WS), boron nitride (BN), rocksalt (RS), distorted rocksalt (d-RS), and distorted wurtsalt (d-WS). At $T_\mathrm{eff}$ = 0 K (i.e., perfect cation order and no disorder), the WS phase is computed to be the most stable phase. This is the phase that had been predicted to be a p-type TC. However, the key takeaway of this plot is that as disorder increases, the WS phase becomes destabilized and instead, RS becomes the most stable phase. This supports experimental results, where RS is synthesized rather than the target WS, and future work is needed to explore whether WS can actually be synthesized. These results are from the author's previous work on \ce{ZnZrN2},\cite{woods2022role} however we expect that this mechanism may present in other predicted p-type TCs in which the ordered 0 K ground state has not yet been synthesized.

\subsubsection{\ce{Ta2SnO6}}

The representative plot for \ce{Ta2SnO6} from Barone et al. depicts that although a region of chemical potential space exists in which \ce{Ta2SnO6} is "hole dopable" (green arch), this is not the region that is sampled experimentally.\cite{barone2022growth} This region is identified as a function on a phase stability diagram as a function of temperature and oxygen partial pressure. In the pink region, which is probed by MBE synthesis, oxygen vacancies are high and limit the Fermi level of \ce{Ta2SnO6} such that p-type doping is compensated by "hole killers." In the white region, which is sampled via annealing, the \ce{Ta2SnO6} is unstable and decomposes. This calculation shows that although \ce{Ta2SnO6} has not yet been grown highly p-type, theoretically such a dopable region should exist, and perhaps follow-up experimental work can target synthesis within this region. We believe that studies such as this one, using computational methods to identify target regions of chemical potential space as a function of experimentally-achievable variables, could be very useful in other chemical systems to accelerate experimental growth of predicted p-type TCs.

\subsection{Defect calculations}

\begin{figure*}
    \centering
    \includegraphics[width=120mm]{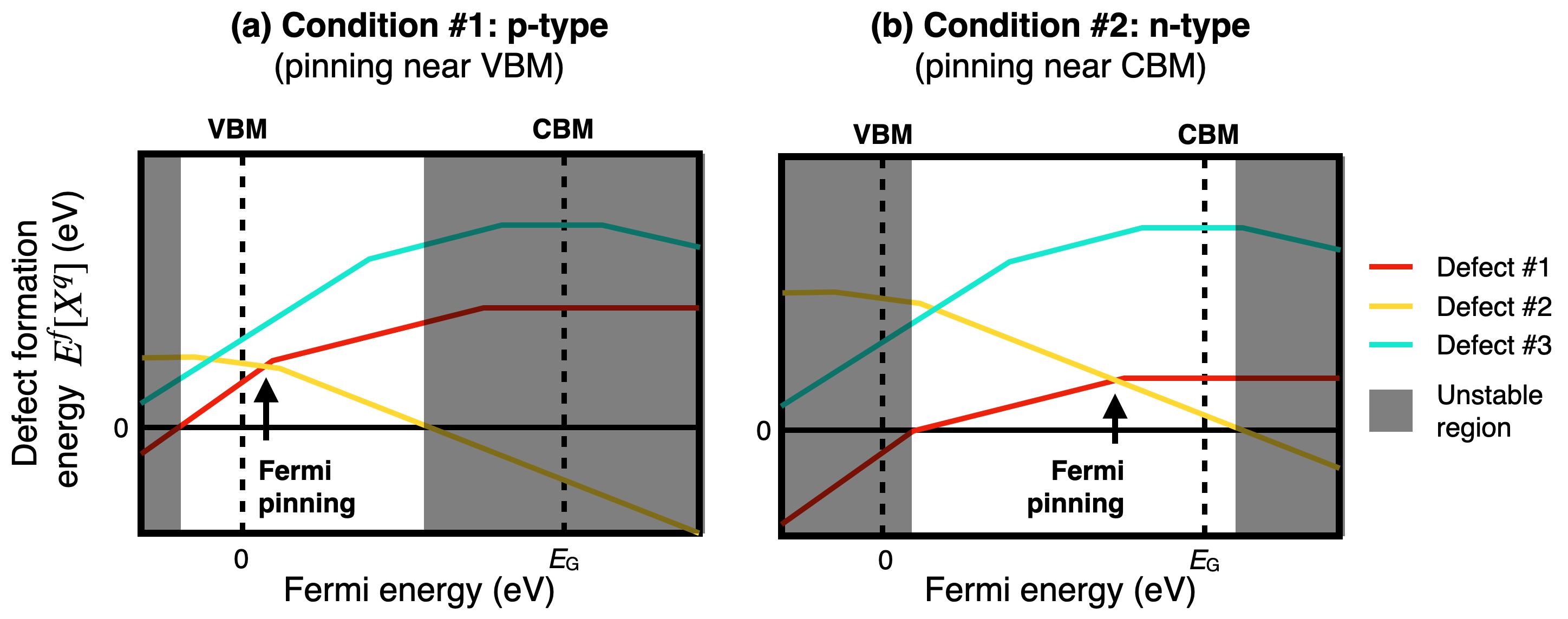}
    \caption{Schematic of various defect formation energy diagrams in which the Fermi level is "pinned" (a) near the VBM (p-type) and (b) near the CBM (somewhat n-type). Figure is reproduced with permission from Woods-Robinson et al., 2021.\cite{woods2021bridging}}
    \label{fig:defect-schematic}
\end{figure*}

The representative plots for \ce{Cs4CdSb2Cl12} and \ce{Ba2BiTaO6} are defect formation energy diagrams. A simple schematic for the interpretation of these diagrams is shown in \autoref{fig:defect-schematic}, reproduced with permission from the lead author's dissertation. The valence band minima (VBM) and conduction band maxima (CBM) are depicted as dashed lines, and three hypothetical defects are represented (\#1, \#2, \#3), and the Fermi level is "pinned" at the point of lowest energy in which a positive-sloped (i.e., positively-charged) defect crosses a negative-sloped (i.e., negatively-charged) defect. The distance of the pinning from the VBM determines the degree of p-type doping such that in (a) a somewhat p-type dopable region is accessible whereas in (b) p-type doping is unlikely to occur. We note that each facet of chemical potential space requires a unique defect formation energy calculation, and these are the underlying calculations that have determined the "hole-dopable" region in the \ce{Ta2SnO6} figure. For more information on defect formation energy calculations, we refer the reader to comprehensive reviews\cite{vandewalle2004firstprinciples, freysoldt2014firstprinciples, broberg2023high} and python packages.\cite{broberg2018pycdt, goyal2017computational, toriyama2023vtandem, kavanagh2024doped}

\subsubsection{\ce{Ba2BiTaO6}}

\ce{Ba2BiTaO6} is an example of a predicted p-type TC that was synthesized before defect calculations were performed. Early synthesis studies in 2015 yielded challenges with high p-type doping, but it wasn't until 2020 that Dahliah et al. reported hybrid defect formation energy calculations to confirm that indeed p-type doping is likely limited by hole-killers: both oxygen vacancies (V\textsuperscript{O}) and Ta-on-Bi antisites (Ta\textsuperscript{Bi}).\cite{dahliah2020defect} The formation energy literature plot depicts six different defects participating, however our diagram is simplified to only include the three lowest energy defects (V\textsuperscript{O}, Ta\textsuperscript{Bi}, and K\textsuperscript{Ba}). We note that in the same year Yin et al. used PBE0 to also confirm Ta\textsuperscript{Bi} as the primary compensating donor, and suggest future work to synthesize \ce{Ba2BiTaO6} under Ta-poor conditions.\cite{yin2020double} This is an example of why defect calculations are always useful to include during the computational screening phases whenever possible.

\subsubsection{\ce{Cs4CdSb2Cl12}}

The \ce{Cs4CdSb2Cl12} from Hu et al. depicts various defects within a formation energy diagram in the highest dopable region of chemical potential space, computed both by PBE and HSE functionals.\cite{hu2019first} Our simplified schematic of this figure (on the right) includes only the lowest energy defects --- cadmium interstitials (Cd\textsuperscript{i}) and Cd-on-Sb antisites Cd\textsuperscript{Sb} --- since these are the defects that determine dopability conditions. Notably, although earlier studies using PBE calculations predicted high hole dopability (p-type with no Fermi pinning observed very close to the VBM), higher order theory using HSE calculations told a different story. With HSE, the Cd\textsuperscript{i} defect is lower in energy than that computed with HSE, and this serves to compensate hole doping and limit the Fermi level to much lower doping levels. This study showcases the importance of using higher levels of theory than PBE to confirm dopability before going to the effort to synthesize and dope a computationally predicted p-type TC; we recommend future studies to always do this.

\subsection{Absorption}

Lastly, \ce{CaCuP} and \ce{Cr2MnO4} are selected to demonstrate absorption-related challenges.

\subsubsection{CaCuP}

For \ce{CaCuP}, which was first predicted by Williamson et al. to be transparent,\cite{williamson2016engineering} Willis et al. showed that experimentally \ce{CaCuP} is much less transparent than expected.\cite{willis2022prediction} This is explained in part by absorption at high wavelengths associated with phonon-mediated indirect transitions, computed by accounting for electron-phonon ("e-ph") coupling. However, this only partially explains the reduction of transparency, and the authors propose that the discrepancy between experiment (blue curve) and e-ph calculations (red curve) may be due to imperfections in crystallinity and extended defects. This study highlights a computational disconnect, and warns researchers that absorption calculations using only the independent particle approximation (in which phonons and defects are not accounted for) may in many cases highly overestimate transparency. 

\subsubsection{Cr2MnO4}

\ce{Cr2MnO4} represents the classic trade-off in TC research: increasing doping can significantly reduce transparency. The figure selected from Nagaraja et al. showed that although high transparency has been observed in undoped and nondegenerate-doped \ce{Cr2MnO4},\cite{peng2013li} high hole dopings and alloying of \ce{Cr2MnO4} with Li significantly increase absorption.\cite{nagaraja2014experimental} Future work to estimate this trade-off before synthesis could be useful for accelerating realization of predicted p-type TC.

\section{Additional supplemental material}

Additional supplemental material includes a data table with the following:

\begin{itemize}
    \item Computational predictions of p-type TCs.
    \item Experimental reports of predicted p-type TCs.
    \item Parsed literature review by material.
    \item Benchmarking of computational vs. experimental reports.
    \item Literature reports of valence and conduction band offsets in p-type TCs.
    \item Reported performance, properties, and metadata of perovskite solar cells with p-type TCs used as contacts, parsed from the Perovskite Database.\cite{jacobsson2022open}
\end{itemize}

\noindent A series of \texttt{jupyter} notebooks to generate Figure 1(c), Figure 6, and SI figures is included. Selected supplemental data is also available via on the MPContribs platform.

\bibliographystyle{ieeetr}
\bibliography{main.bib}